\documentclass[english,prd,aps,10pt,twocolumn,floatfix,nofootinbib,superscriptaddress]{revtex4-1}
\usepackage{lipsum}
\usepackage{float}
\usepackage[utf8x]{inputenc}
\usepackage{textcomp}
\usepackage{amsmath}
\usepackage{amsfonts}
\usepackage{amssymb}
\usepackage{graphicx}
\usepackage{hyperref}
\usepackage{verbatim}
\usepackage{subfigure}
\usepackage[usenames, dvipsnames]{color}

\newcommand\insertimg[5] 	  
{\begin{figure}[!htbp]							
    \begin{center}							
      \includegraphics[width=#2cm, height=#3cm]{#1}
	    \caption{#4}\label{#5}
    \end{center}
\end{figure}}
									
\newcommand\insertfig[4]  	 
{\begin{figure}[!htbp]					
    \begin{center}						
	    \includegraphics[scale=#2]{#1}		
	    \caption{#3}\label{#4}				
    \end{center}
\end{figure}} 

\begin{document}

\title{Astrophysical jets from boosted compact objects} 

\author{Ramiro Cayuso}
\email{rcayuso@perimeterinstitute.ca }
\affiliation{Facultad de Matem\'{a}tica, Astronom\'{i}a, F\'{i}sica y Computaci\'{o}n, Universidad Nacional de C\'o{}rdoba.\\
Instituto de F\'{i}sica Enrique Gaviola, CONICET. Ciudad Universitaria (5000), C\'o{}rdoba, Argentina. }
\affiliation{Perimeter Institute for Theoretical Physics, Waterloo, ON, N2L2Y5, Canada.}
\author{Federico Carrasco}
\email{federico.carrasco@uib.es}
\affiliation{Facultad de Matem\'{a}tica, Astronom\'{i}a, F\'{i}sica y Computaci\'{o}n, Universidad Nacional de C\'o{}rdoba.\\
Instituto de F\'{i}sica Enrique Gaviola, CONICET. Ciudad Universitaria (5000), C\'o{}rdoba, Argentina. }
\affiliation{Departament de F\'i{}sica \& IAC3, Universitat de les Illes Balears and Institut d'Estudis Espacials de Catalunya, Palma de Mallorca, Baleares E-07122, Spain.}
\author{Barbara Sbarato}
\email{barbarasbarato@gmail.com}
\affiliation{Facultad de Matem\'{a}tica, Astronom\'{i}a, F\'{i}sica y Computaci\'{o}n, Universidad Nacional de C\'o{}rdoba.\\
Instituto de F\'{i}sica Enrique Gaviola, CONICET. Ciudad Universitaria (5000), C\'o{}rdoba, Argentina. }
\author{Oscar Reula}
\email{reula@famaf.unc.edu.ar}
\affiliation{Facultad de Matem\'{a}tica, Astronom\'{i}a, F\'{i}sica y Computaci\'{o}n, Universidad Nacional de C\'o{}rdoba.\\
Instituto de F\'{i}sica Enrique Gaviola, CONICET. Ciudad Universitaria (5000), C\'o{}rdoba, Argentina. }
%

\date{\today}

\begin{abstract}
We perform full 3D numerical simulations of compact objects, such as black holes or 
neutron stars, boosted through an ambient force-free plasma that posses a uniform magnetization. 
We study jet formation and energy extraction from the resulting stationary late time solutions.
The implementation of appropriate boundary conditions has allowed us to explore a wide range of boost velocities, finding the jet power scales as $\gamma v^2$ (being $\gamma$ the Lorentz factor). 
We also explore other parameters of the problem like the orientation of the motion respect to the asymptotic magnetic field or the inclusion of black hole spin. 
 Additionally, by comparing a black hole with a perfectly conducting sphere in flat spacetime, we manage to disentangle curvature effects from those produced by the perfect conducting surface. It is shown that when the stellar compactness is increased these two effects act in combination, further enhancing the luminosity produced by the neutron star.
 

\end{abstract}


\maketitle

\section{Introduction}

Enormous amounts of energy, in the form of Poynting winds or highly collimated relativistic jets, are often observed in various astrophysical scenarios. Such energetic phenomena are believed to be powered by compact objects like black holes (BH) and neutron stars (NS), from the interactions with strong and large-scale magnetic fields on their surrounding magnetospheres.
In the seminal works of Goldreich \& Julian \cite{goldreich} and Blandford \& Znajek \cite{Blandford} (describing pulsars and active galactic nuclei, respectively), it was first demonstrated that the vicinity of these spinning compact objects would be filled with a tenuous plasma. In such rarefied environments, the electromagnetic force dominates over particle inertia and leads to a great simplification in the problem, 
allowing to capture the basic mechanism that taps rotational energy by means of the electromagnetic field.
While pulsars admits a classical interpretation as Faraday disks \cite{faraday1832}, in the black hole scenario the energy is instead extracted in a form of generalized Penrose process (see e.g.~\cite{lasota2014}) known as the Blandford-Znajek mechanism. This low-inertia limit of relativistic magnetohydrodynamics, referred as force-free electrodynamics (FFE), has been --since then-- widely used to study global properties of neutron stars and black holes magnetospheres, like for instance Refs.~\cite{contopoulos1999,  Komissarov2004b, mckinney2006relativistic, timokhin2006force, spitkovsky2006, Palenzuela2010Mag, Gralla2014}.

In the force-free approximation, when there is a perturbation of an otherwise constant magnetic field, the dynamics makes these perturbation travel preponderantly along the magnetic field lines, thus carrying energy with them along this direction. In this work, we simulate a couple of astrophysically relevant situations where this happens, which consist on a black hole or a neutron star moving through a plasma-filled region of constant magnetic field.
Galactic mergers could provide a likely scenario for the black hole case \cite{begelman1980,milosavljevic2005}, since the resulting circumbinary disk of the merged galaxy will anchor magnetic field lines, some of which traverse the central region where the binary --and eventually the final supermassive black hole-- moves. 
Another example could be a BH-NS binary, in which the black hole would move through the magnetic field of a neutron star \cite{mcwilliams2011,paschalidis2013}.
In such cases, we expect the black hole to loose some kinetic energy, transforming it by enlarging its mass but also into electromagnetic energy propagated away by the jets. There has been a number of previous numerical studies on this scenario, \cite{palenzuela2010dual, Palenzuela2010Mag, Luis2011}, which we use as starting point for the present work. All of them analyze the problem from the point of view of the stationary magnetic field, namely, in their numerical grid the black hole moves and creates the jets which carry the energy. The advantage is that they can readily compute an approximate -since their time direction is a not Killing direction for the background geometry-  energy flux. It is precisely this absence of a timelike Killing vector field and the correspondingly lack of a conserved positive-definite energy, which permits to have energy transport via jets in this approximation where the background is fixed. The disadvantage is that they can not model high speed black holes for they move too quickly outside the grid. 
In our case we choose to describe the problem from the black hole static geometry. The black hole sees a boosted magnetic field and the corresponding electric field, the interaction of its geometry with that electromagnetic configuration generates a stationary solution which takes energy away through jets.  In our case we do have a background Killing vector field, and so conservation of the energy it defines, but this is not the energy an observer for which the (uniform) magnetic field is at rest would see. Thus, we also have to define approximate energy fluxes corresponding to these --for our description-- moving observers. The energies so defined are transported away, as expected.

The other situation we model is that of a neutron star, also moving on a region of uniformly magnetized plasma. This could happen if a neutron star orbits near an active supermassive black hole, where both strong magnetic fields and force-free plasma are expected around the central region. It could also be relevant in the context of electromagnetic precursor signals from neutron stars mergers \cite{palenzuela2013electromagnetic, palenzuela2013linking, ponce2014}, the likely progenitors of gamma-ray burst. 
We consider here an idealized setting where the neutron star is represented by a perfectly conducting spherical surface and there is no field generated at the stellar interior. This might be regarded as the limiting case in which the exterior magnetic field is much stronger than the one associated to the star, so that the later can be neglected. We defer the inclusion of the star's own magnetic field and rotation to a more detailed analysis on a future work. 
A similar behavior to the boosted (nonspinning) black hole case is found, although the details of the operating mechanisms are not the same. Here, kinetic energy from the motion is transformed into Alfvén waves, sourced by the boundary conditions at the conductor. One nice aspect of this problem is that it allows to take the flat spacetime limit, in which the boosted time direction is also a Killing direction. This means there is no ambiguity on defining the energy fluxes used for the description; and thus, it might help gaining some insight into the previous --more delicate-- black hole scenario.


In section II, we describe the setup for both cases: the numerical scheme, geometry and evolution equations; the initial data, the boundary conditions and, finally, the energy fluxes definitions. 
With all of these information one should be able to reproduce our results unambiguously. Except for the boundary conditions and energy fluxes, the setting is very similar to the one in \cite{palenzuela2010dual, Luis2011}. 
In section III we present the results of our simulations, where different aspects of the problem were explored. Conclusions and perspectives are drawn on Section IV. 
Throughout, we adopt geometrized units in which $c=G=1$ and Lorentz-Heaviside units for the electromagnetic field.

\section{Setup}

We are interested on modeling the magnetosphere of a compact object (BH or NS) that travels across a uniform magnetic field by solving the equations of force-free electrodynamics. 
The code used here was first described in \cite{FFE2} for black holes and later extended in \cite{NS}, by developing appropriate boundary conditions for the perfectly conducting surface of a neutron star.
Since we adopt the reference frame of the central object, its motion relative to the uniform magnetic field will be accomplished through suitable boundary conditions at the external surface of the domain.
We shall look for stationary solutions obtained by evolving the fields until they do not change appreciably. 
The resulting state is determined only by boundary conditions, the background geometry, and to some extent on the handling of the electric field growth on the current sheets that the dynamics generates.

Although a detailed description of our numerical implementation can be found on previous works \cite{FFE2,NS}, we shall start this section by briefly summarizing its basic features along with information about the metric and the set of evolution equations employed. Then, we shall describe initial data and boundary conditions for the two scenarios we want to study. And finally, we shall discuss the energy fluxes definitions used to analyze the results. 

\subsection{Numerical Implementation}

We evolve a particular version of force-free electrodynamics derived at \cite{FFE}, which has some improved properties in terms of well posedness and involves the full force-free current density~\footnote{Similar hyperbolic formulations were presented in Refs.~\cite{pfeiffer,Pfeiffer2015}.}. More concretely, we shall consider the evolution system given by Eqs.~(8)-(9)-(10) in \cite{NS}.
The numerical scheme to solve these equations is based on the \textit{multi-block approach} \cite{Leco_1, Carpenter1994, Carpenter1999, Carpenter2001}, in which the numerical domain is built from several non-overlapping grids where only grid-points at their boundaries are sheared.
The equations are discretized at each individual subdomain by using difference operators constructed to satisfy summation by parts (SBP). 
In particular, we employ difference operators which are sixth-order accurate on the interior and third-order at the boundaries. 
Numerical dissipation is incorporated through the use of adapted Kreiss-Oliger operators. These compatible difference and dissipation operators were both taken from Ref.~\cite{Tiglio2007}.
\textit{Penalty terms} \cite{Carpenter1994, Carpenter1999, Carpenter2001} are added to the evolution equations at boundary points.
These terms penalize possible mismatches between the different values the characteristic fields take at the interfaces, 
providing a consistent way of communicate information between the different blocks:
essentially, the outgoing characteristic modes of one grid are matched onto the ingoing modes of its neighboring grids.
At each subdomain, it is possible to find a semi-discrete energy defined by both a symmetrizer of the system at the continuum and a discrete scalar product (with respect to which SBP holds). 
The summation by parts property of the operators allows one to obtain an energy estimate, up to outer boundary and interface terms left after SBP. 
The penalties are constructed so that they make a contribution to the energy estimate which cancels inconvenient interface terms, thus providing an energy estimate which covers the whole integration region across grids. 
Such semi-discrete energy estimates --provided an appropriate time integrator is chosen-- guarantee the stability of the numerical scheme \cite{Kreiss, Carpenter1994, Carpenter1999, Carpenter2001, Leco_2}. 
A classical fourth order Runge-Kutta method is used for time integration in our code. 



We use a particular multiple patch infrastructure that has been equipped with the Kerr metric, as in Ref.~\cite{Leco_1}. This provides a numerical domain that is perfectly adapted to the geometry of the problems, having two global inner/outer boundaries with spherical topology~\footnote{See figure 2 of \cite{FFE2} for an illustration of the numerical domain.}. 
The Kerr metric is parametrized by the mass $M$ and spin $a$, and  can be written in the Kerr-Schild form as $g_{ab} = \eta_{a b} + H \, \ell_{a } \ell_{b } $, where $\eta_{ab}$ is the flat metric and $\ell_{a}$ is a null co-vector with respect to both $\eta_{ab}$ and $g_{ab}$. For visual representation, throughout this article, we will present our results in the Cartesian coordinates $\{t,x,y,z\}$ 
associated with the flat part of the metric\footnote{Sometimes referred as the Kerr-Schild Cartesian coordinates, or the Kerr-Schild frame (see e.g.~\cite{Visser2007}).}. 
In these coordinates, the metric function $H$ takes the form 
\begin{eqnarray}
H &=& \frac{2 M r}{r^2 + a^2 z^2 /r^2}\\
r^2 &=& \frac{1}{2}(\rho ^2-a^2) + \sqrt{\frac{1}{4}(\rho ^2-a^2)^2 + a^2 z^2} \\
\rho^2 &=& x^2+y^2+z^2  
\end{eqnarray}
and the co-vector $\ell_{a}$ reads
\begin{equation}
 \ell_{a} = \left\lbrace 1, \frac{rx + ay}{r^2+a^2}, \frac{ry-ax}{r^2+a^2}, \frac{z}{r} \right\rbrace .
\end{equation}

Typically we solve in a region between an interior sphere whose radio is either inside the black hole or represents a perfectly conducting boundary, and an exterior spherical surface at $r=162M$. This region is covered by a total of $9\times 6$ grids, being $9$ the number of layers. The typical resolution used on each of these grids is of $41 \times 41$ grid-points in the angular directions and $N_r =101$ points for the radial one. The grids layers do not cover regions of identical radial extension, having more resolution near the inner boundary than in the asymptotic region: from layer to layer we decrease the effective radial resolution by a factor $1.3$. 
In some cases, we have increased the resolution of the individual grids to $61 \times 61$ and $N_r =151$. 

In order to handle current sheets, we use a rather standard approach in which electric field is effectively dissipated to maintain the condition that the plasma is magnetically dominated (i.e.~$B^2 -E^2 >0$),  discussed in \cite{FFE2}. At the current sheets  the magnetic field presents a jump discontinuity in the vertical direction in the $y$ component while the electric field has a spike in the $x$ component. 
When high order finite difference operators are used in such discontinuous regions, the fields behave in an unsatisfactory manner.  
Indeed, high order operators tend to give a noisier results. To overcome this issue, the precision of the finite difference operators is reduced from 6th to 2nd order for those grids covering the region where current sheets form. 
Thus, providing a substantial improvement in the quality of the numerical approximation. 

\subsection{Initial Data}

\subsubsection{Boosted Black Hole}

We consider a black hole moving with velocity $v_{o}$ respect to a reference frame in which the magnetic field is asymptotically uniform and along the $z$-axis, while the electric field vanishes.
The boost direction is not necessarily orthogonal to this magnetic field, so we shall study the evolution of the system for different alignments between the black hole velocity and the asymptotic magnetic field. 
Since we adopt the reference frame in which the black hole is at rest, the electric and magnetic fields should arise from a Lorentz transformation of the electromagnetic field from the frame in which 
the magnetic field is uniform and the electric field vanishes. That is, 
\begin{equation}\label{B_prime}
\vec{B}=\gamma  \, \vec{B}^{\prime} - \frac{\gamma^{2}}{\gamma + 1}(\vec{v} \cdot \vec{B}^{\prime}) \, \vec{v}
\end{equation}
\begin{equation}\label{E_prime}
\vec{E}= \gamma(\vec{v}\times \vec{B}^{\prime}).
\end{equation}
where $\gamma=\frac{1}{\sqrt{1-v^{2}}}$ is the Lorentz factor.
 In Kerr-Schild Cartesian coordinates, the uniform magnetic field would read
\begin{equation}
B^{\prime x}=B^{\prime y}=0 \quad , \quad B^{\prime z}= \frac{B_{0}}{\sqrt{h}}
\end{equation}
while the velocity can be generically written as
\begin{equation}\label{velocity}
v^{x}= 0 , \quad v^{y}=v_{0}\cos(\chi), \quad v^{z}=v_{0}\sin(\chi)
\end{equation}
being $\chi$ the angle among the velocity and the $y$-axis.

%
We emphasize the fact that the steady state solutions would only depend on the boundary conditions, namely on the asymptotic boosted fields, and not on the particular way we have chosen to set the initial configuration in the interior.\\


\subsubsection{Boosted perfectly conducting sphere}

We shall consider a perfectly conducting surface as an idealized model of a neutron star. In the present work, we have assumed that the interior magnetic field of the star is several orders of magnitude weaker than the exterior uniform magnetic field. Thus, we propose an initial data that comes from the configuration of an asymptotically uniform magnetic field around a superconducting sphere, which ensures a vanishing normal component of the magnetic field at the stellar surface. 
Concretely, the initial condition is given by 
\begin{equation}\label{ini_cond_conf}
\vec{B} = B_{0} \hat{z} - \frac{B_{0}R^{3}}{2r^{3}}\left( \frac{3z}{r}\hat{r} - \hat{z}\right) 
\end{equation}
\\
where $R$ is the radius of the star. Hence the field is tangent to the stellar surface and asymptotically matches a uniform magnetic field along the $z$-axis, as we wanted. Notice, however,  that it does not satisfies the magnetically dominated plasma condition at the poles where the magnetic field vanishes. Since the force-free equations would break down at these points, we simply chose a grid that does not contain them, which has proven to be enough for achieving well-behaved solutions.

\subsection{Boundary Conditions}

As mentioned before, our numerical domain is bounded by two inner/outer spherical surfaces, where boundary conditions needs to be specified. The treatment given in the present article to these boundaries was previously described and employed in Refs.~\cite{FFE2,NS}.
However, we find important to briefly summarize here the main aspects\footnote{We refer to \cite{FFE2,NS} for a complete discussion and technical details.} and explain how these boundary conditions are applied in this new astrophysical context.
Generally speaking, physical conditions are imposed by fixing appropriately  all the incoming characteristic (physical) modes via the penalty method. 
Whereas for the constraints, in this case the divergence-free condition $\nabla \cdot \vec{B}=0$, we adopt a method presented in \cite{FFE2} (see also \cite{Mari}) which restricts possible incoming violations at both boundary surfaces.

Our implementation of the outer boundary conditions consist on setting the incoming physical modes according to a fixed source $U_{ext} = (\vec{E}_{ext}, \vec{B}_{ext})$ that we control. This idea appear motivated on the interfaces treatment and was already employed in Ref.~\cite{FFE2}, where $U_{ext}$ represented a uniform magnetic field sourced by a distant accretion disk.
We shall use this strategy again here, with $U_{ext}$ now being the boosted electromagnetic configurations that threatens the compact object magnetosphere. 
Thus, for the black hole case, $U_{ext}$ is given by the boosted uniform magnetic field (see eqns. \eqref{B_prime}-\eqref{velocity}). While for the neutron star, the source is given by the boost of its initial configuration \eqref{ini_cond_conf} evaluated at the boundary. In this later case, we shall introduce the boost smoothly to its final boost velocity $v_0$ (at time $t_f$) by using a time-dependent function
\begin{equation}
 \frac{v_0}{2} \left[ 1 - \cos (\pi t / t_f ) \right] \quad \text{if} \quad 0 \leq t \leq t_f
\end{equation}

The treatment of the inner boundary is, on the other hand, very different between the black hole and neutron star scenarios. In the first case, the inner edge of the domain is simply placed inside the black hole horizon where all characteristic modes points inward (i.e. they are all outgoing from our numerical domain perspective), and hence, there are no incoming modes to be prescribed. 
In the neutron star case, the inner edge of our domain is placed at the stellar surface which is assumed to behave as an idealized perfect conductor. 
In the present work, we have further assumed that the interior magnetic field of the star is negligibly small with respect to the one of the external magnetized plasma and also that the star is not rotating. Thus, the boundary conditions reduces to
\begin{equation}
B^r = 0 , \quad \alpha E_{\theta} = \sqrt{h} \beta^r B^{\phi}, \quad \alpha E_{\phi} = -\sqrt{h} \beta^r B^{\theta}
\end{equation}
with $\alpha$, $\beta^i$ and $h_{ij}$ being the \textit{lapse function}, \textit{shift vector} and the \textit{intrinsic metric} on the spatial slices, respectively.
The normal magnetic field is keep fixed to zero by enforcing it at each Runge-Kutta substep, as done in Ref.~\cite{PHAEDRA}.
While the electric field components are imposed through the \textit{penalty method}, by fixing the incoming physical modes to a rather involved combination of outgoing modes.
We refer the interest reader to \cite{NS} (in particular Sec.~II-C and the Appendix) for further details on this.

\subsection{Fluxes}

In this section we shall discuss the relevance of different quantities needed to describe the jets and the energy extraction process.
Before computing any quantity it is important to recall that the electromagnetic field does not by itself define any four-momentum, a four-vector and the energy-momentum tensor are necessary to build this quantity. The different choices of that four-vector define different four-momenta that can be thought to be related to different observers. Thus, the result obtained by computing the electromagnetic flux (as an spatial component of the four-momentum) in the BH's frame will certainly be different from the one obtained in the plasma frame of reference. 
The electromagnetic flux expression for the BH's frame comes naturally, it is associated to the time-like Killing vector field of the background spacetime geometry that allows to construct a conserved four-momentum current.
To obtain fluxes this vector is contracted to the normal to some space-like hypersurface, usually the boundary of a sphere at some radius $r=R$ in Kerr-Schild coordinates. 
We will refer to this quantity as Poynting flux, defined as follows,
\begin{equation}\label{flux_pf}
  \Phi_{ \mathcal{E}}:= \sqrt{-g} \, p^r
\end{equation}
where $p^{a}$ is the four-momentum defined by,
\begin{equation}\label{cuadri-momento}
  p^{a} := - T^{ab}k_{b}
\end{equation}
where $T^{ab}$ is the Energy-momentum tensor and $k^{a}$ is the Killing vector field related to stationarity (see for instance Appendix B of \cite{FFE2}).
The factor $\sqrt{-g}$ appears from the normalization at the surface $r=R$.

Now the task is to find a quantity that is representative of the electromagnetic energy flux in the reference frame in which the BH is not static, namely the frame where the asymptotic magnetic field is constant and at rest. To represent an observer that moves relative to the BH with velocity $v^{\prime a}$, we can take its four-velocity $n^{\prime a}$ to be,

\begin{equation}\label{tprima}
    n^{\prime a} = \gamma ( n^{a} + v^{\prime a})
  \end{equation}
One can thus define the four-momentum $p^{\prime a}$ for this observer as,
\begin{equation}\label{cuadri-momento_prima}
  p^{\prime a} := - T^{ab}n^{\prime}_{b}  \quad 
\end{equation}
where $n^{a}$ is the normal vector to the equal-time hypersurface $\Sigma_{t} $ of the space-time foliation.

Since the background geometry describes a curved spacetime, this boosted quantity is rather arbitrary: it does not have the same normalization as the Killing vector field nor gives a conserved current. Nevertheless it acquires meaning as an asymptotic quantity, far away from the BH we can use the fact that the geometry is approximately flat there and so the boosted time direction would approach a boosted Killing vector of the underlying asymptotic geometry. 
Thus, for large distances we can define the four-momentum of an observer that is in the plasma rest frame (i.e. with velocity $v^{\prime a} = - v^{a}$) as in \eqref{cuadri-momento_prima}.

Now using this four-momentum we can define a quantity that is representative of the electromagnetic flux in this frame, we can do so by contracting $p^{\prime a}$ with a vector $N^{\prime a}$ that is both of norm unity and orthogonal to $n^{\prime a}$. 
By proposing an \textit{Ansatz} of the form $N^{\prime a} = ( N^{a} + \zeta v^{\prime a} + \xi k^{a} )$ (where $k^{a}$ is the Killing vector field associated with stationarity and $N^{a}$ is the radial unit vector) and using the conditions:

\begin{equation}\label{conditions}
  N^{\prime a}N^{\prime}_{a} = 1  \quad ; \quad  n^{\prime a}N^{\prime}_{a} = 0
\end{equation}

we can obtain the value of the constants $\xi$ and $\zeta$ as, 
\begin{equation}\label{a_m2}
  \xi = -\frac{N_{\beta}}{\alpha} + v^{\prime r} + \zeta v^{\prime 2}{\alpha - v^{\prime}_{\beta}}  
\end{equation}
%
\begin{equation}\label{b_m2}
    \begin{split}
    \zeta = & \frac{\sqrt{\alpha ^{2}(\alpha - v^{\prime}_{\beta})^{2}((v^{\prime r}\alpha^{2} -  \beta ^{r}v^{\prime 2}_{\beta} )^{2}  + v^{\prime 2}q(v^{\prime r 2}\alpha ^{2} - \beta ^{r 2} ) )}}{\alpha ^{2} v^{\prime 2}(\alpha ^{2} + v^{\prime 2}q - v^{\prime 2}_{\beta})}\\
 & + \frac{ (\alpha - v^{\prime}_{\beta})(\beta ^{r}\alpha v^{\prime}_{\beta} - v^{\prime r}\alpha^{3}) + v^{\prime 2}q \alpha ( \beta ^{r} - \alpha v^{\prime r}) }{\alpha ^{2} v^{\prime 2}(\alpha ^{2} + v^{\prime 2}q - v^{\prime 2}_{\beta})}
    \end{split} 
 \end{equation}
where $q = (-\alpha ^{2} + \beta ^{2}) $, $ v^{\prime}_{\beta} = v^{\prime i} \beta _{i}$ and $\alpha$ $\&$ $\beta$ are respectively the Lapse and Shift of the spacetime foliation. 
It can be easily checked that for large values of r (i.e. where $\alpha \rightarrow 1$ and $\beta \rightarrow 0$) the vector $N^{\prime a}$ is simply the Lorentz transformation of the $N^{a}$ vector, and so the surface determined by the vectors $N^{\prime a}$ is asymptotically a boosted sphere.

Now we can finally define the electromagnetic energy flux for the boosted frame as 
\begin{equation}\label{PF_boost}
\Phi^{\prime}_{ \mathcal{E}}:= \sqrt{h }p^{\prime a}N^{\prime}_{a}
\end{equation}
where $h$ is the determinant of induced metric to the surface in which this flux is computed.

It is very important to stress that, even though the expression \eqref{PF_boost} can be evaluated in the whole numerical domain, it's value will only be representative of the electromagnetic flux far away from the BH, where $n^{\prime a}$ approaches an asymptotic Killing vector field and consequently the $p^{\prime a}$ is an asymptotically conserved four-momentum. 
This downside of not being able to properly define the electromagnetic flux (for the plasma's reference frame) in the whole numerical domain is not a consequence of our choice of the BH rest reference frame. 
In previous works, \cite{Luis2011,alic2012,palenzuela2010dual} a measurement of the electromagnetic flux is also given accurately only far away from the Black Hole(s) system since the Killing field needed to construct the conserved four momentum  is only an asymptotic concept.

Another aspect that has to be taken into account is the fact that actually this system is not isolated, since uniform magnetic and electric fields are present as a background in the whole numerical domain and they would remain so as a consequence of the boundary conditions imposed at the outer boundary. 
Special care has to be taken in order to distinguish the electromagnetic flow generated by the BH's interaction with the fields, from the ubiquitous flow generated by the background. In order to subtract adequately this background radiation we take the same approach as in \cite{Luis2011,alic2012}, that is, we subtract the value of the field's initial condition to the stationary values of the electromagnetic field (final configuration) before computing the value of the electromagnetic energy flux.
Another interesting approach to this problem is to compute asymptotic fluxes using the plane wave structure of force-free electrodynamics, that is, its fast and Alfvén propagation modes. Mensurable quantities can be constructed for each plane wave mode and study them separately. For instance, we shall look at the radial fluxes $\Phi^{\pm}_A$ (Alfvén modes) and  $\Phi^{\pm}_T$ (fast magnetosonic) of the final stationary solutions. Following appendix A of Ref.~\cite{NS} and assuming our solutions reasonably satisfy the constraint, i.e.  $\vec{E}\cdot \vec{B} \approx 0$, we get:
\begin{eqnarray}
 \Phi^{\pm}_A &:=& \lambda^{\pm}_A  (\Theta^{\pm}_A (U))^2 (u^{\pm}_A)^2 \nonumber \\
 &=& (\beta_m + \alpha \sigma^{\pm}_A ) \frac{E_{m}^2}{\left[ 1-(\sigma^{\pm}_A)^2 \right] } \label{A-flux}\\
 \Phi^{\pm}_T &:=& \lambda^{\pm}_T  (\Theta^{\pm}_T (U))^2 (u^{\pm}_T)^2 \nonumber\\
 &=& (\beta_m \pm \alpha ) \frac{\left[ B_{p}^2 - E^2 \right] ^2}{\left[ B_{p}^2 + E_{p}^2 \mp 2 S_m \right] } \label{F-flux}
\end{eqnarray}
where $\sigma_{A}^{\pm} := \frac{1}{B^2} \left( S_m \pm \sqrt{B^2 - E^2} B_m \right) $, $A_m$ represents contractions of vectors on the radial unitary direction $m^i$ and, $A_{p}^i := A^i - A_m m^i$, its perpendicular projections.


\section{Numerical Results}

\subsection{Boosted black hole}

In this section we shall explore different parameters of the problem such as boost velocity $v$, black hole spin $a$ and inclination angle $\chi$ among the boost direction and the $y$-axis (i.e.  $\chi$ represents the departure of the direction of the motion from the case in which it is orthogonal to the asymptotic magnetic field).
%
Lets start first by the case where the asymptotic magnetic field is orthogonal to the boost velocity $v^{a}$ (e.i. $\chi=0$). We shall study some aspects of the solution for different magnitudes of this velocity, such as the topology of the electric and magnetic field configurations, the power of the electromagnetic flux and the development of a current sheet. 
Later, we shall study the dependence of the luminosity on the misalignment between the boost direction and the asymptotic magnetic field.
And finally, we will analyze the effects that including rotation has on the electromagnetic flux. 

\subsubsection{Orthogonal velocity}

\begin{figure}
  \begin{center}
\includegraphics[scale=0.23]{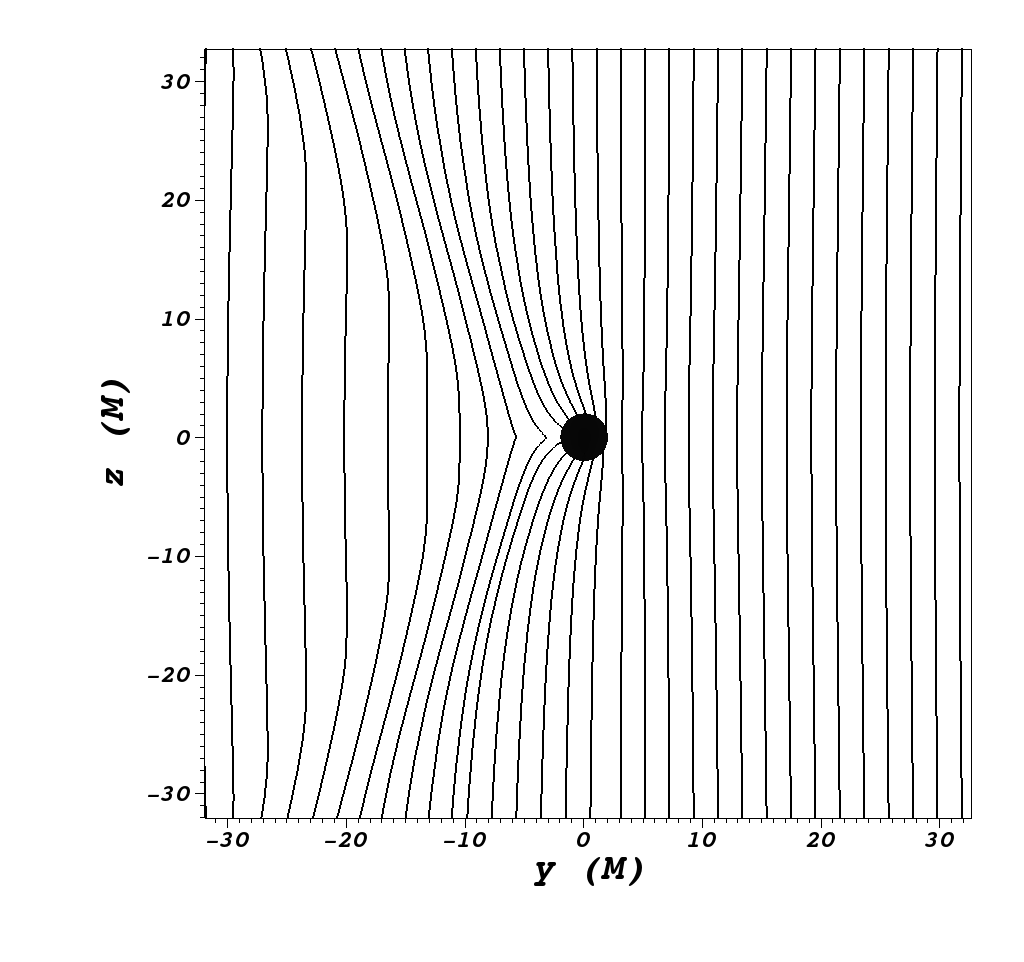}
\includegraphics[scale=0.23]{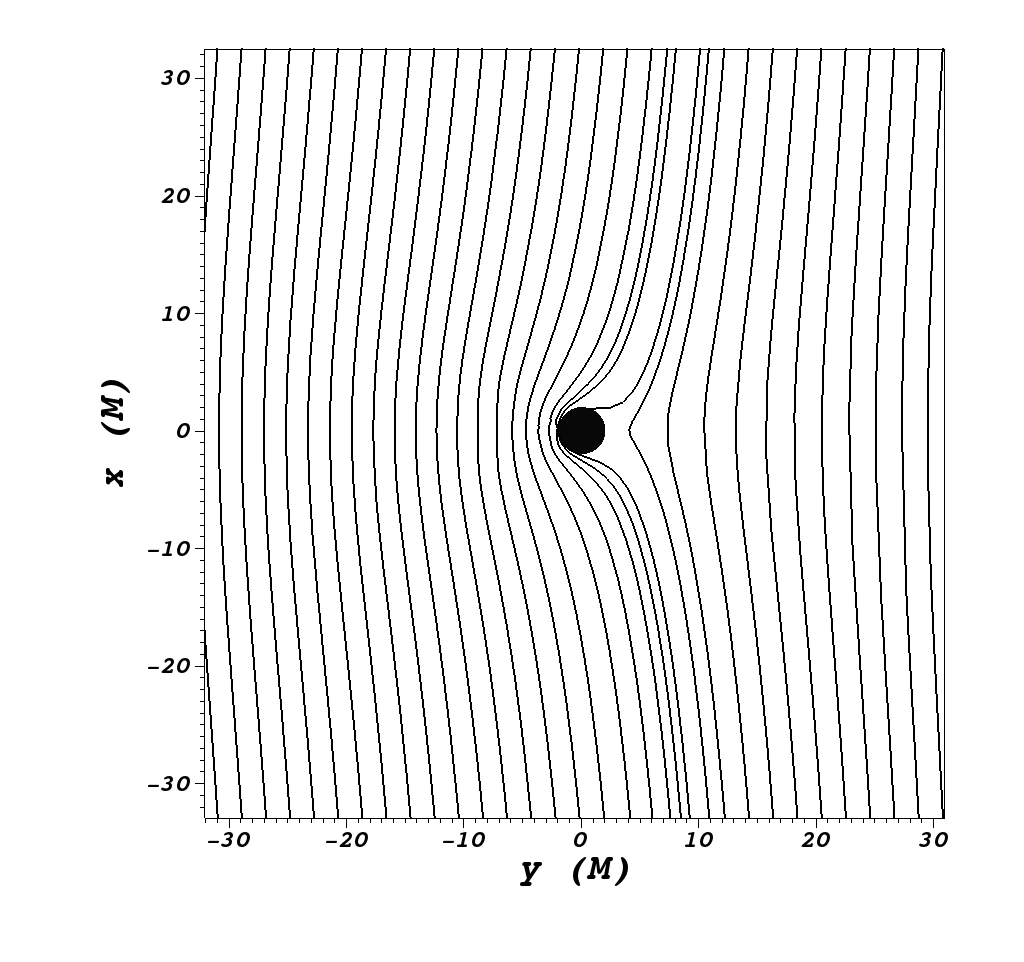}
  \caption{ Late time numerical solution ($t = 400M)$ for a black hole moving at speed $v=0.5$ along the $y$-direction. Streamlines of the magnetic field at the $x=0$ plane (top) and of the electric field at the $z=0$ plane (bottom) are illustrated. }
 \label{fig:fieldlines}  
 \end{center}
\end{figure}

\begin{figure}
  \begin{center}
\includegraphics[scale=0.26]{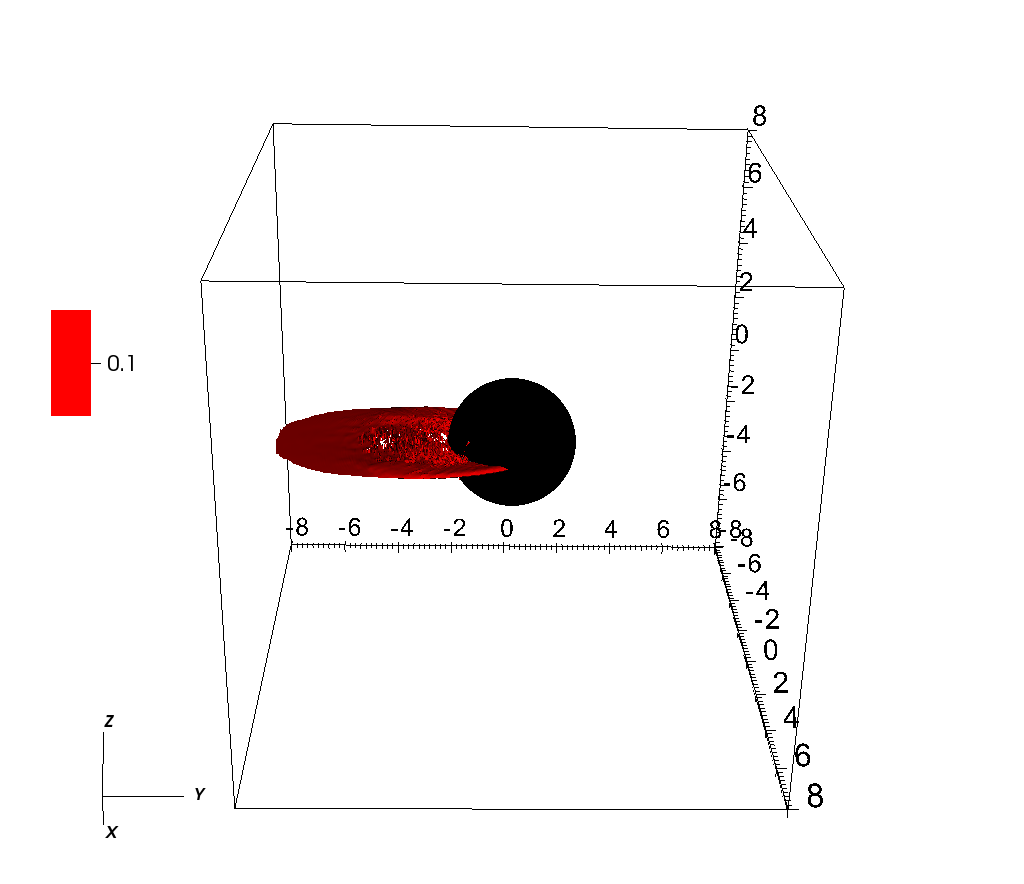}
  \caption{Surface where $\frac{B^{2} - E^{2}}{B^{2}}=0.1 $, for a late time solution of a black hole moving at speed $v=0.5$ along the $y$-direction. It signals the presence of a strong current sheet, where electric fields is effectively dissipated.}
 \label{Current_sheet} 
 \end{center}
\end{figure}

We present a late time configuration for the case of a Schwarzschild black hole, boosted with velocity $ v=0.5$ orthogonal to the asymptotic magnetic field. The general structure of the solution is depicted on Fig.~\ref{fig:fieldlines}, where it can be seen that magnetic field lines are disturbed by the passage of the black hole, leaving a trail behind it. Similarly, electric field lines, shown at the $z=0$ plane, are  also dragged by the black hole as it moves along the $y$-axis, but from the opposite side. The way in which the magnetic field is pulled towards the black hole, result in a discontinuity of the $y$-component at the equatorial plane, thus producing a current sheet with strong electric fields.
In Fig.~\ref{Current_sheet} we have plotted the quantity $\frac{B^{2}-E^{2}}{B^{2}}$, which when close to zero signals the location this current sheet. 
In such regions, the numerical mechanism that effectively dissipates electric field is actively operating to avoid violations of the magnetic domination condition. We see that for the present case, the current sheet extends behind the black hole's motion, up to approximately $6M$. 

%
\begin{figure}[t!]
  \begin{center}
  \begin{minipage}{8cm}
\subfigure{\includegraphics[scale=0.23]{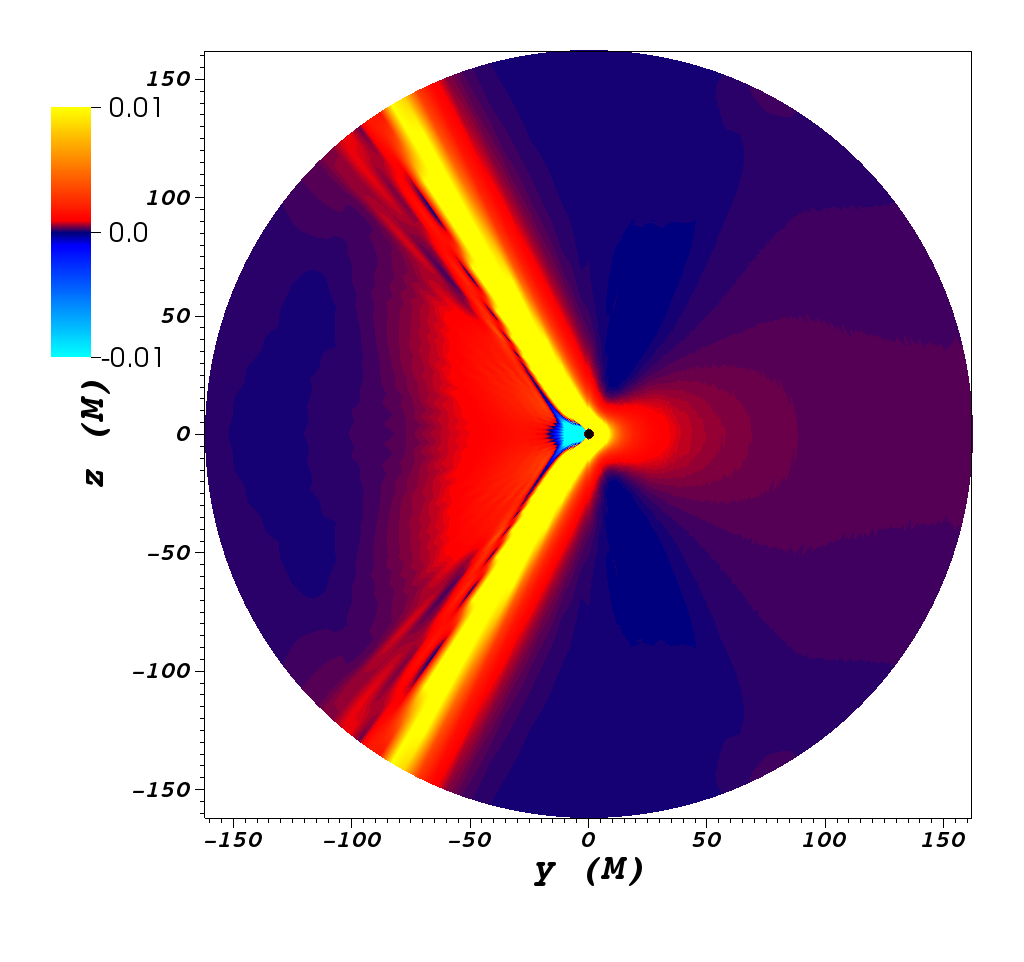}}
\end{minipage}
\begin{minipage}{8cm}
\subfigure{\includegraphics[scale=0.23]{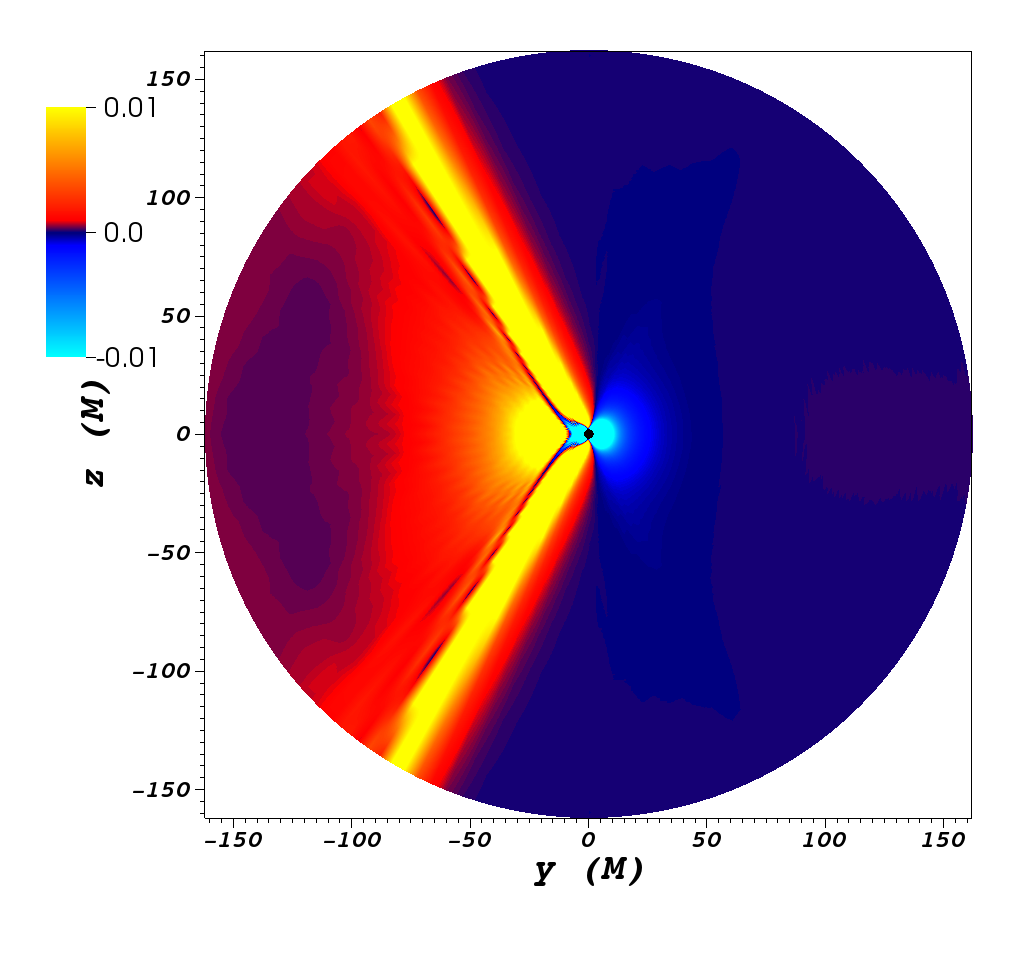}}
\end{minipage}
  \caption{Comparison of the electromagnetic flux densities in the  $x=0$ plane for the orthogonal boost velocity $v=0.5$. 
   \textbf{Top:} Poynting flux density, $p^{\prime a}N^{\prime}_{a}$,  measured by an observer at rest with the asymptotic uniform magnetic field.
   \textbf{Bottom:} Poynting flux density,  $p^{r}$, as measured in the black hole frame.}
 \label{pfn_v05} 
 \end{center}
\end{figure}

Figure \ref{pfn_v05} displays the radial electromagnetic energy flux density on the $x=0$ plane, as measured in the two relevant reference frames of the problem, namely: the one where the observer is at rest with respect to the asymptotic magnetic field (plasma frame, top image) and the one in which the black hole is at rest (BH frame, bottom image). Both the flux in the plasma frame, i.e. $p^{\prime a}N^{\prime}_{a}$, and the flux measured in the BH frame, $p^{r}$, exhibit a pair of highly collimated jets emerging from the black hole. These jets form an angle with the $z$-axis given by, $\theta_{jet} = \tan^{-1} (\gamma v)$, in the co-moving frame; and equivalently, $\theta_{jet}^{\prime} = \tan^{-1} (v)$, in the plasma frame. Such misalignment between the collimated energy flux and the original magnetic field orientation is expected and has been reported previously in \cite{Luis2011}. At a first glance, we see the main difference between the two fluxes is at their non-collimated components, especially in front of the black hole where it gives negatives values in the BH frame.

\begin{figure}
  \begin{center}
\includegraphics[scale=0.23]{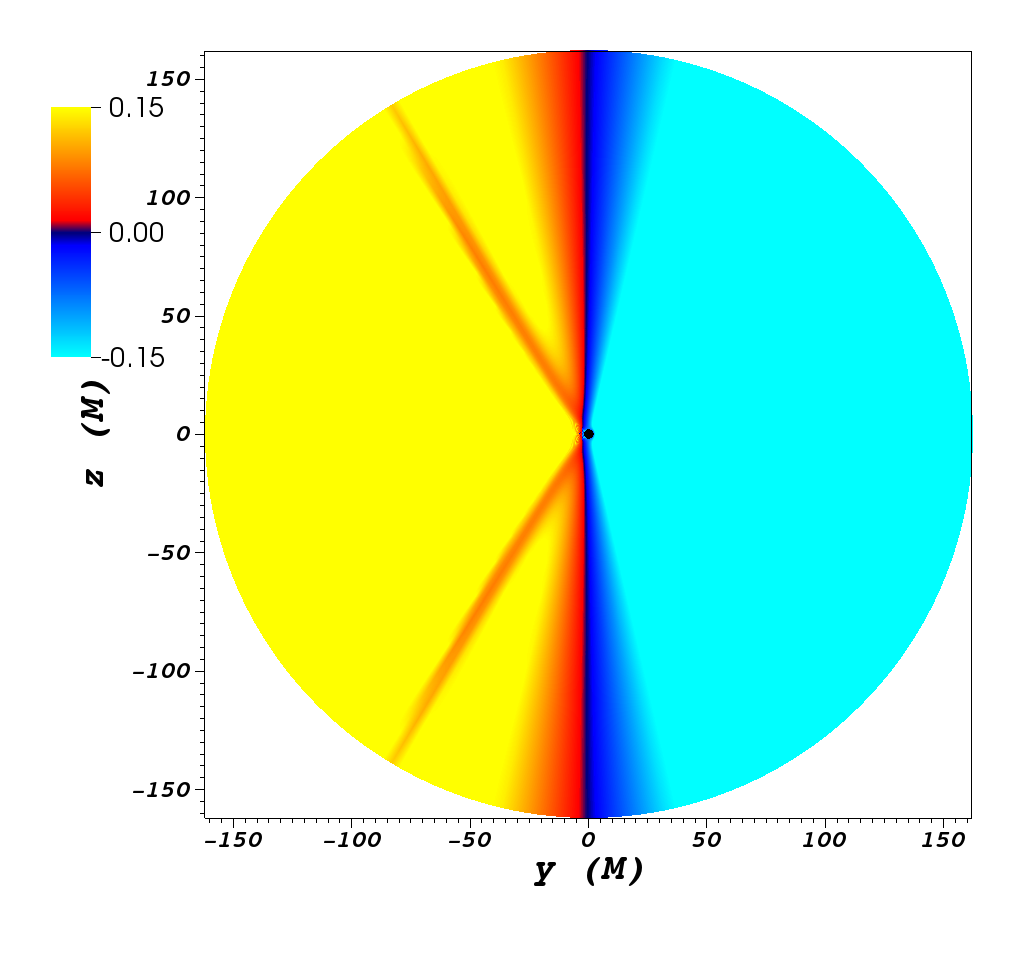}
  \caption{Electromagnetic flux density  $p^{r}$ in the $x=0$ plane for the orthogonal boost velocity $v=0.5$, as measured in the black hole frame when the background initial field is not subtracted from the final stationary field configuration.} 
 \label{PFN_back} 
 \end{center}
\end{figure}

\begin{figure}
  \begin{center}
\includegraphics[scale=0.3]{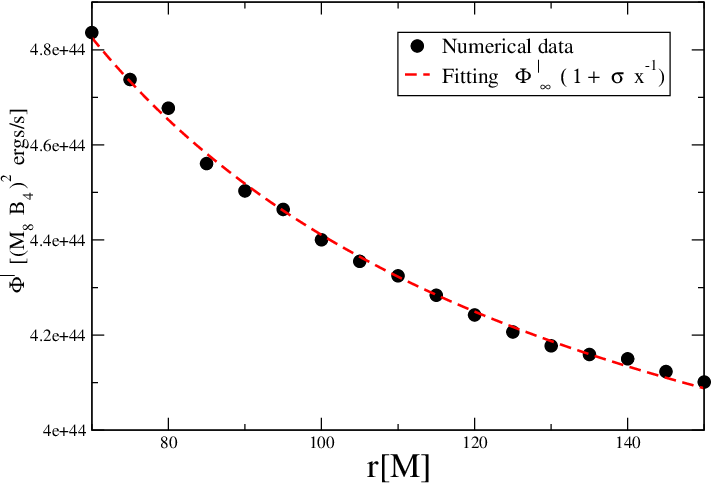}
  \caption{Dependence of the net collimated flux $\Phi^{\prime}$ with integration radius $r$, for the black hole moving at speed $v=0.5$.}
 \label{R_dependence} 
 \end{center}
\end{figure}

Figure \ref{PFN_back} presents again the radial electromagnetic energy flux density as measured in the BH's frame, but now the initial background fields has not been subtracted from the final stationary configuration. The same pair of jets as in Fig.~\ref{pfn_v05} can be seen, except that they are somewhat hidden now by a mainly dipolar flux density distribution arising from the background electromagnetic field being boosted against the black hole. It is worth emphasizing that integrating this flux around the BH horizon gives a rather small but negative value, thus showing that there is no energy extraction from the black hole. This is consistent with the fact that there is no ergoregion here and, hence, the Blandford-Znajek mechanism is not possible. The net positive flow of electromagnetic energy in the plasma frame, on the other hand, must arise from the available energy due to the relative motion between the magnetized plasma and the black hole.
We turn next to a more quantitative analysis and consider, first, how does the emitted jet power changes with the distance from the black hole horizon. Thus, we measure the integrated flux in the plasma frame, $\Phi^{\prime}$, as a function of  radius. The integration is performed on a surface determined by the normal vector $N^{\prime a}$ (that  in this frame represents spheres), within a cylindrical region of  $60M$ diameter enclosing the collimated jet.
Following the notation of Refs.~\cite{palenzuela2010dual,Luis2011}, we shall compute this quantity in physical units, respect to a representative system in which a black hole of mass $M=10^8 M_{\odot}$ is immersed on a magnetic field of strength $B_o = 10^{4} G$. That is, the results will be expressed proportional to $(M_{8} B_{4})^2 \equiv (\frac{M}{10^8 M_{\odot}})^2  (\frac{B}{10^4 G})^2$, allowing for an easy translation to any pair of physical values $M$ and $B$.
Figure \ref{R_dependence} presents the behavior of $\Phi^{\prime}$ in the range $r=70-150 M$, for a black hole moving at speed $v=0.5$. It can be seen that the emitted power drops approximately $20\%$ between $r=70M$ and $r=150M$. A function of the form $\Phi^{\prime} \sim \Phi_{\infty}(1 + \sigma r^{-1})$, also shown in the plot, fits the numerical data very well. The asymptotic value for the collimated flux is $\Phi_{\infty} = 3.44 \times 10^{44} \textit{erg} / s$, and $\sigma = 28.1 M$.  
The expression $\Phi^{\prime}$ we propose to compute this flux is an approximation that relies on an asymptotic Killing vector field and is only a local integration. Hence, it is prone to errors and should be considered only as a guidance. Too close to the BH the uncertainties on the approximation to the asymptotic Killing vector field are important and, far away, there are dispersion effects. Thus, we fixed an intermediate radius $r=90M$ as the integration surface to measure the collimated energy flux $\Phi^{\prime}$.

\begin{figure}[t!]
  \begin{center}
  \begin{minipage}{8cm}
 \subfigure{\includegraphics[scale=0.3]{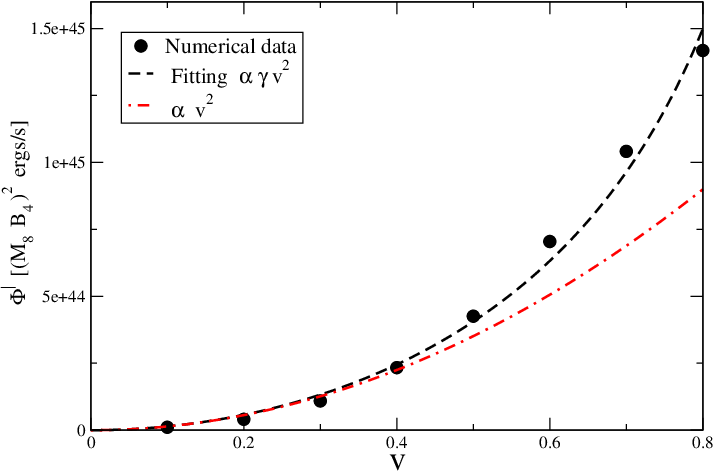}} 
\end{minipage}
\begin{minipage}{8cm}
\subfigure{\includegraphics[scale=0.3]{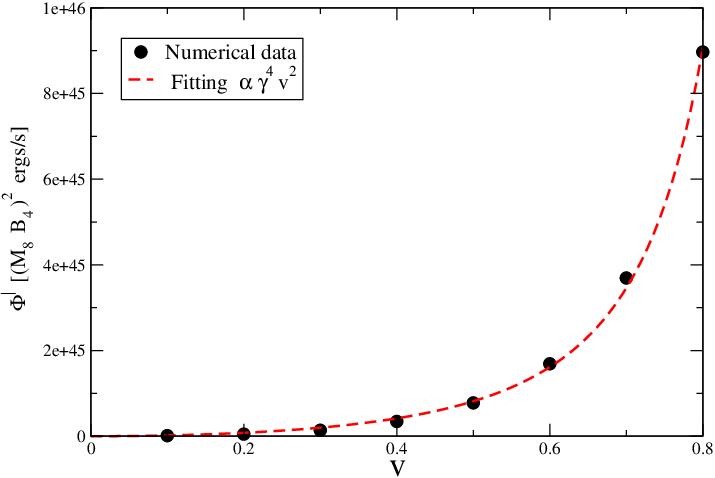}}
\end{minipage}
  \caption{Comparison of the dependence on the black hole velocity for the total  EM collimated flux for a black hole traveling in the $y$. 
   \textbf{Top:} Total integrated EM collimated flux, $p^{\prime a}N^{\prime}_{a}$,  measured by an observer at rest with the asymptotic uniform magnetic field. The dots correspond to the numerical values, while the red and black curves are fittings of the form  $\Phi^{\prime}$ $\propto$ $ \gamma v^{2}$ and $\Phi^{\prime}$ $\propto$ $ v^{2}$ .
   \textbf{Bottom:} Total integrated EM collimated flux, $p^{r}$, as measured in the black hole frame. The dots correspond to the numerical values, while the red curve is a fitting of the form  $\Phi$ $\propto$ $ \gamma ^{4} v^{2}$.}
 \label{pf_v} 
 \end{center}
\end{figure}

The results obtained for several boost velocities for both reference frames are summarized on Fig.\ref{pf_v}. For the plasma frame (top figure), as expected from previous numerical experiments  in the regime $v \leq 0.2$ \cite{Luis2011}, further supported on theoretical arguments \cite{morozova2014, penna2015} later, the emitted power for non-relativistic speeds goes as $\propto v^{2}$ (see red dashed curve).  
However, we find that for larger boost velocities the correct dependence is instead given by $\propto \gamma v^{2}$ (solid black line), which fits the numerical values very well for the whole range of velocities explored. To the best of our knowledge, no one has pointed out this behavior before, which may have important observational consequences on astrophysical scenarios were such relativistic speeds are plausible. Meanwhile, from the BH's frame, the behavior with the boost velocity is instead given by $\propto \gamma ^{4} v^{2}$. 

\subsubsection{Misaligned case}

Now, we consider situations in which the exterior magnetic field and the black hole velocity are not orthogonal. 
In order to do that, we take the boost velocity to lay on the $y-z$ plane and parametrize different orientations by the angle $\chi$ it forms with the $y$-axis (see expression \eqref{velocity}). For this particular scenario, we have focused in the case of a black hole moving at speed $v=0.5$.

\begin{figure}
  \begin{center}
\includegraphics[scale=0.23]{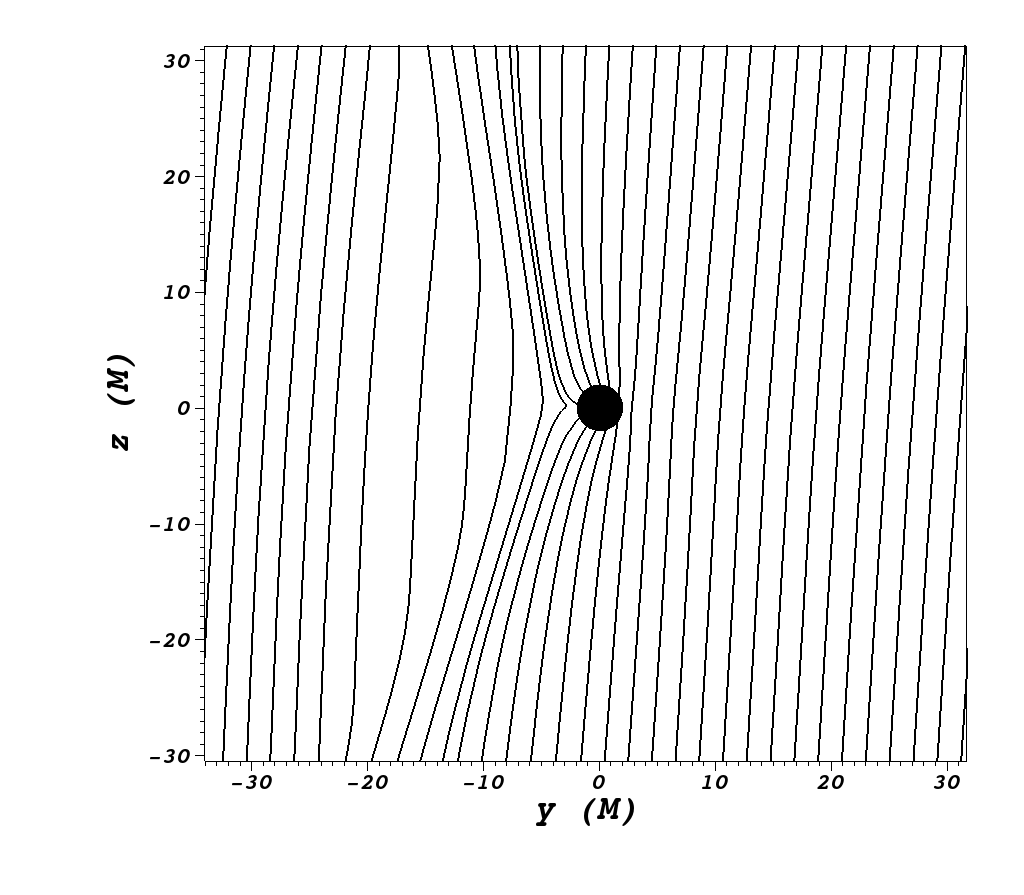}
  \caption{Representative streamlines of the magnetic field for the $\chi=-\pi / 4$ case for the late time numerical solution (t=400M) for a black hole with $v=0.5$.}
 \label{tilted_streamlines} 
 \end{center}
\end{figure}

A representative late time configuration of the magnetic field is shown in Fig.~\ref{tilted_streamlines}, corresponding to a black hole traveling with inclination angle $\chi=-\pi / 4$. 
Notice that the field topology is similar to the one of the orthogonal case (shown in Fig.~\ref{fig:fieldlines}), but now the asymptotic field is rotated an angle $  \Theta = \tan^{-1} (\frac{B^{\prime y}}{B^{\prime z}}) = \tan^{-1} (\frac{v^2 \cos \chi \sin\chi}{\sqrt{1-v^2}+1-v^2 \sin^2 \chi})$ within the $y-z$ plane, as seen from the black hole's frame. Such rotation is induced by the Lorentz transformation and can be straightforwardly computed from equation \eqref{B_prime}.

\begin{figure}
  \begin{center}
\includegraphics[scale=0.23]{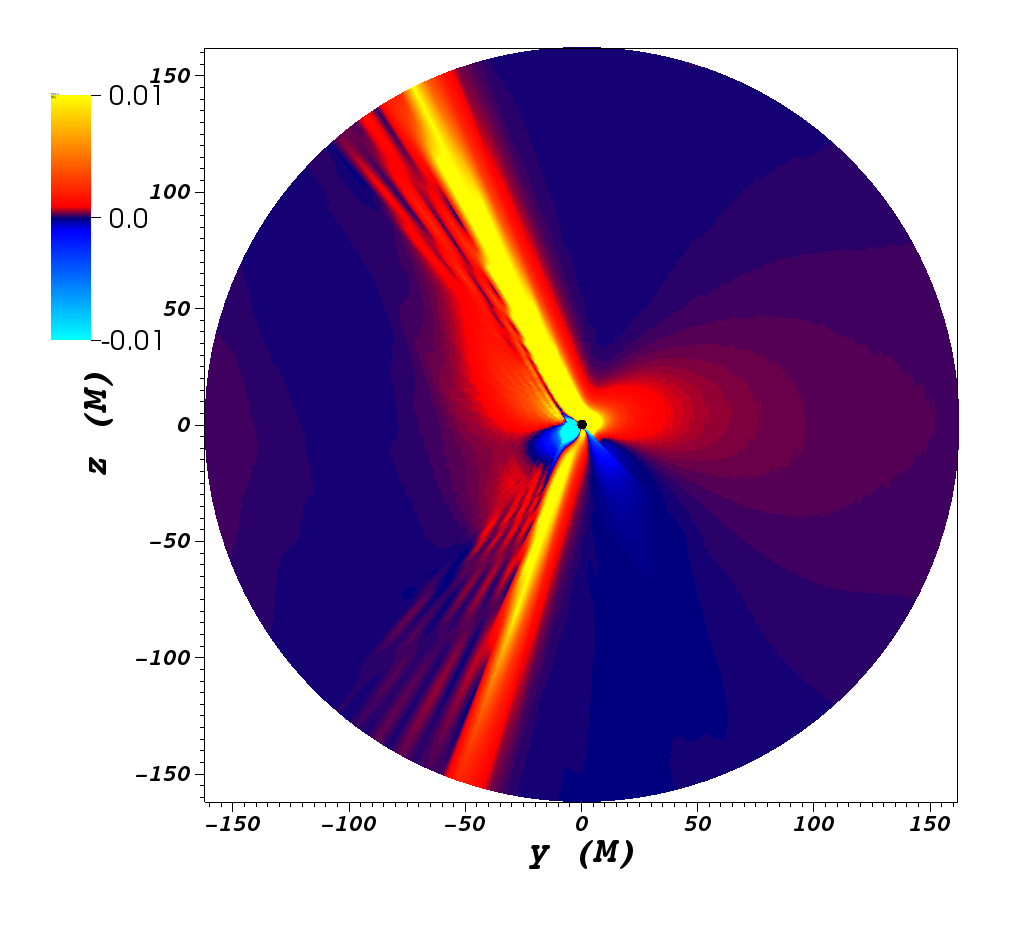}
  \caption{Electromagnetic flux density $p^{\prime a}N^{\prime}_{a}$ for the $x=0$ plane for late time numerical solution ($t = 400M)$ for a black hole with velocity $v=0.5$ and $\chi = \pi / 4$.}
 \label{pfn_over4} 
 \end{center}
\end{figure}

\begin{figure}
  \begin{center}
\includegraphics[scale=0.3]{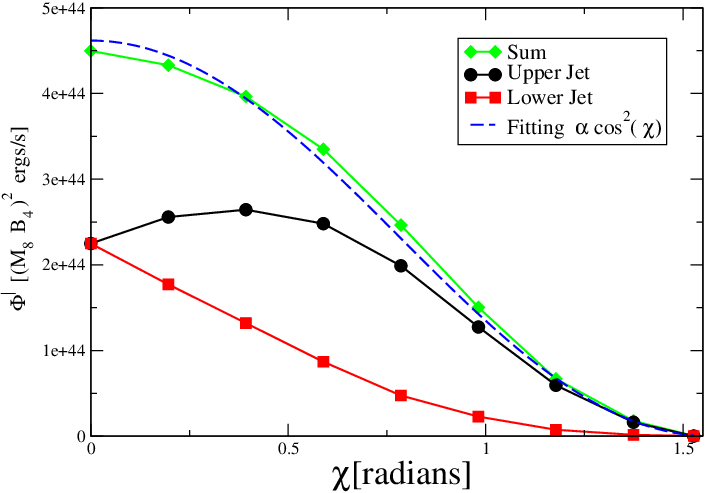}
  \caption{Dependence of the EM flux $\Phi^{\prime}$, for a black hole with $v=0.5$, on the angle $\chi$. The squares (dots) correspond to the net flux integrated near the lower (upper) jet, while the green diamonds correspond to the sum of these quantities. The blue doted curve corresponds to a fitting $\propto \cos^{2}(\chi)$.} 
 \label{angles_pfn} 
 \end{center}
\end{figure}

\begin{figure}
  \begin{center}
\includegraphics[scale=0.3]{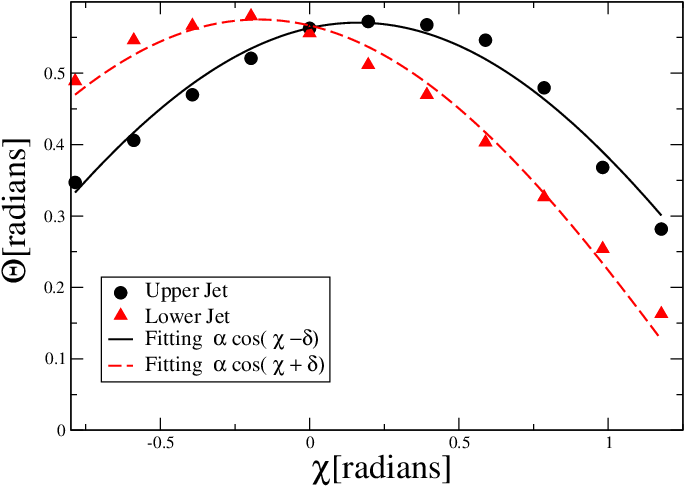}
  \caption{The angle $\Theta$ between the Upper (Lower) jet and the $z$ ($-z$) axis, is given as a function of the inclination angle $\chi$ for a black hole moving at speed $v=0.5$.}
 \label{displacement_angle} 
 \end{center}
\end{figure}

Figure \ref{pfn_over4} shows the EM energy flux density $p^{\prime a}N^{\prime}_{a}$ in the $x=0$ plane, corresponding to a late time solution of a black hole moving with a direction determined by $\chi=\pi/4$. In contrast to Fig.~\ref{pfn_v05}, we see that --as expected-- the solution has lost the reflection symmetry respect to the $z=0$ plane, exhibiting now a pair of asymmetric jets.
To study the dependence of the jet power on inclination, we vary the angle $\chi$ from $0$ to $\pi/2$. It can be seen, in Fig.~\ref{angles_pfn}, that the power for each individual upper/lower jet highly depends on the inclination angle: they can be up to $\approx 17 \%$ higher than in the orthogonal case ($\chi=0$) and vanishing for $\chi = \pi/2$. Figure \ref{angles_pfn} also shows the net collimated power, i.e. the sum of the contributions from both jets, along with a fitting $\propto \cos^{2}(\chi)$, which fits very well with the numerical data. The same $\cos^{2}(\chi)$ behaviour was observed in \cite{FFE2} for the stationary Kerr BH case, with the exception that $\chi$ represents the inclination angle between the asymptotic magnetic field and the BH rotation axis. This behaviour is expected, since the electric field in the BH frame is $\propto \cos (\chi)$.
By performing simulations for negative inclination angles we observed that, as expected, there is a symmetry between the lower and upper jets, in which the upper jet power for a given angle $\chi_o$ equals the power of the lower jet at the opposite angle, i.e. $-\chi_o$.
The simulations performed for different angles $\chi$ also show that the direction of each jet is shifted respect to the $\chi=0$ case, this displacement is shown 
in Fig.\ref{displacement_angle}, which also presents a fitting of the numerical data with functions $\propto \cos(\chi - \delta)$ (for the upper jet) and $\propto \cos(\chi + \delta)$ (for the lower one). For the boost velocity employed here (i.e. $v=0.5$), we find that $\delta = 0.17 \pm 0.01$.

\subsubsection{Spinning black hole}

In this section, we present the results of the simulations performed on a Kerr background, focusing on the effects the black hole rotation has in the emitted power. In Ref.~\cite{morozova2014}, analytic vacuum Maxwell solutions were found for the field configurations in the vicinity of black hole which is both moving and spinning. The estimated luminosities from these solutions have shown that the effect of rotation would be subdominant respect to the one associated with the translation motion. This scenario has also been studied numerically in Ref.~\cite{Luis2011}, within the force-free approximation. Considering boost velocities up to $v=0.2$, the authors of \cite{Luis2011} have proposed a decomposition of the total luminosity as,
\begin{equation}\label{PF_boost_sep}
\Phi = \Phi_{spin} + \Phi_{boost}v^2  
\end{equation}
where $\Phi_{spin}$ and $\Phi_{boost}$ represent the contributions of spin and linear motion, respectively. Thus, suggesting that the two mechanisms acts separately, with $\Phi_{spin}$ being independent of the velocity $v$ (only depending on $a$) and $\Phi_{boost}$ being a constant which does not depend on spin. 

\begin{figure}
  \begin{center}
\includegraphics[scale=0.3]{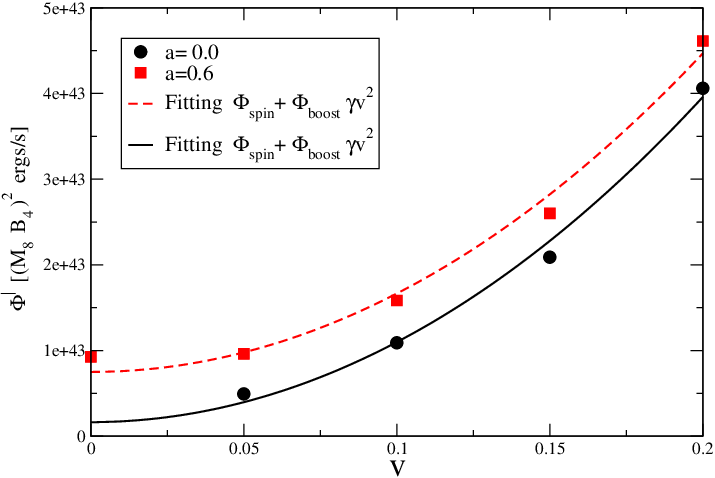}
  \caption{Collimated EM flux $\Phi^{\prime}$ for black holes moving at low velocities, one with $a=0.0$ (black dots) and other with $a=0.6$ (red squares). The curves are fits of the form, $ \Phi_{spin} + \Phi_{boost}\gamma v^2$.}
 \label{a06} 
 \end{center}
\end{figure}

In Fig.~\ref{a06} we show the luminosities at two different spin values: $a=0$ (i.e. non spinning case) and $a=0.6$, for velocities up to $v=0.2$. The plot reproduces very well the results of Ref.~\cite{Luis2011} (specifically, their figure 2), which illustrates --through the fitting curves-- the above mentioned behavior of the two separate contributions. This serves two purposes: it confirm their results by an independent approach to the problem, and on the other hand, it further validates our numerical implementation.

\begin{figure}
  \begin{center}
\includegraphics[scale=0.3]{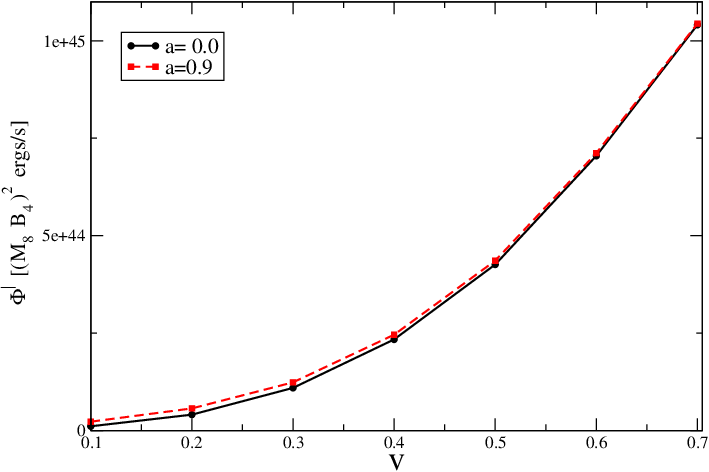}
  \caption{Dependence on the BH's velocity of the total $\Phi^{\prime}$ EM collimated flux for black holes moving with a boost velocity $v$, one with $a=0.0$(red squares) and other with $a=0.9$ (black dots). The two curves approach as the velocity increases.}
 \label{a00_vs_a09} 
 \end{center}
\end{figure}

\begin{figure}
  \begin{center}
\includegraphics[scale=0.3]{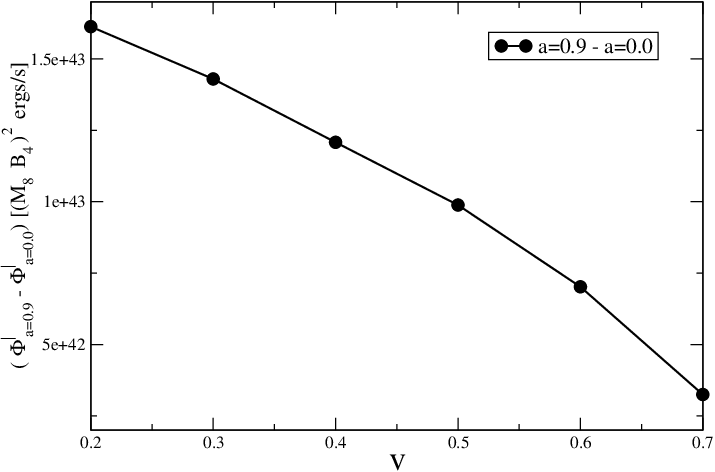}
  \caption{Difference of the two curve from Fig.\ref{a00_vs_a09} e.i. $\Phi^{\prime}_{a=0.9}-\Phi^{\prime}_{a=0.0}$ and its dependence with the boost velocity $v$.}
 \label{a09_minus_a09} 
 \end{center}
\end{figure}

Now, we shall explore what happens if one goes to larger speeds. 
To that end, we present in Fig.~\ref{a00_vs_a09} the resulting luminosities at two spin values, $a=0$ and $a=0.9$, for velocities that ranges from $v=0.1$ to $v=0.7$.
Surprisingly, we find that the curves that represent the non-spinning and the highly-spinning black holes tend to overlap for highly-relativistic boost velocities, $v \gtrsim 0.5$. 
It means the decomposition \eqref{PF_boost_sep} made above no longer holds for such speeds, where
$\Phi_{spin} $ is seen to decrease (see Fig.~\ref{a09_minus_a09}).
This can be explained, at least in part, by noticing that the power on the BZ mechanism diminishes with the inclination angle among the rotation axis and the asymptotic magnetic field. Thus, the angle $\theta_{jet}(v)$ produced by the motion would tend to reduce the spin contribution from the total luminosity. 
Indeed, we have confirmed numerically that by aligning the spin axis to the upper jet (since it can not be aligned simultaneously to both jets) one gets a larger spin contribution $\Phi_{spin} $ than compared to the case in which the rotation axis is perpendicular to the motion and, moreover, that this difference increments with boost velocity. 
\subsection{Boosted perfectly conducting sphere}

We turn now to the idealized setup of a neutron star that is moving across a uniformly magnetized plasma in the orthogonal direction.
The star has been modeled by a perfectly conducting spherical surface on a Schwarzschild spacetime, and it was further assumed to have no magnetic field on its own\footnote{ Physically, it would correspond to the limit in which the exterior magnetic field totally overwhelms the internal field of the star.}.  The star is smoothly brought to relative motion by gradually boosting the initial electromagnetic configuration at the outer boundary. After an initial dynamical transient, the numerical solutions reach a steady state showing collimated electromagnetic jets. We will analyze these solutions and how their jet power vary with the boost velocity $v$ and stellar compactness $\mathcal{C} \equiv M/R$.
Of particular interest is the flat spacetime limit, $M=0$, for that in such case the existence of the Killing vectors makes the notion of the boost and fluxes well defined everywhere, and thus allows for an interesting comparison with the previous black hole scenario.

In a force-free environment, conductors are shown to act as sources by imposing boundary conditions on the surrounding fields \cite{gralla2016}.  A similar setting to our star embedded in flat spacetime has been studied in the context of satellites (see e.g.~\cite{1965drag}), where the problem is fairly well understood. The motion ($\hat{y}$) across a uniform magnetic field ($\hat{z}$) induces charge separation along the transverse ($\hat{x}$) direction, which is conducted away through the plasma in the form of Alfvén waves. An stationary electric circuit, as the one depicted on Fig.~\ref{fig:circuit}, is then established. Such configurations gives rise to a  significant damping on the motion of the object, as mechanical energy is converted to Alfvén radiation \cite{1965drag}.

\begin{figure}[H]
  \begin{center}
\includegraphics[scale=0.23]{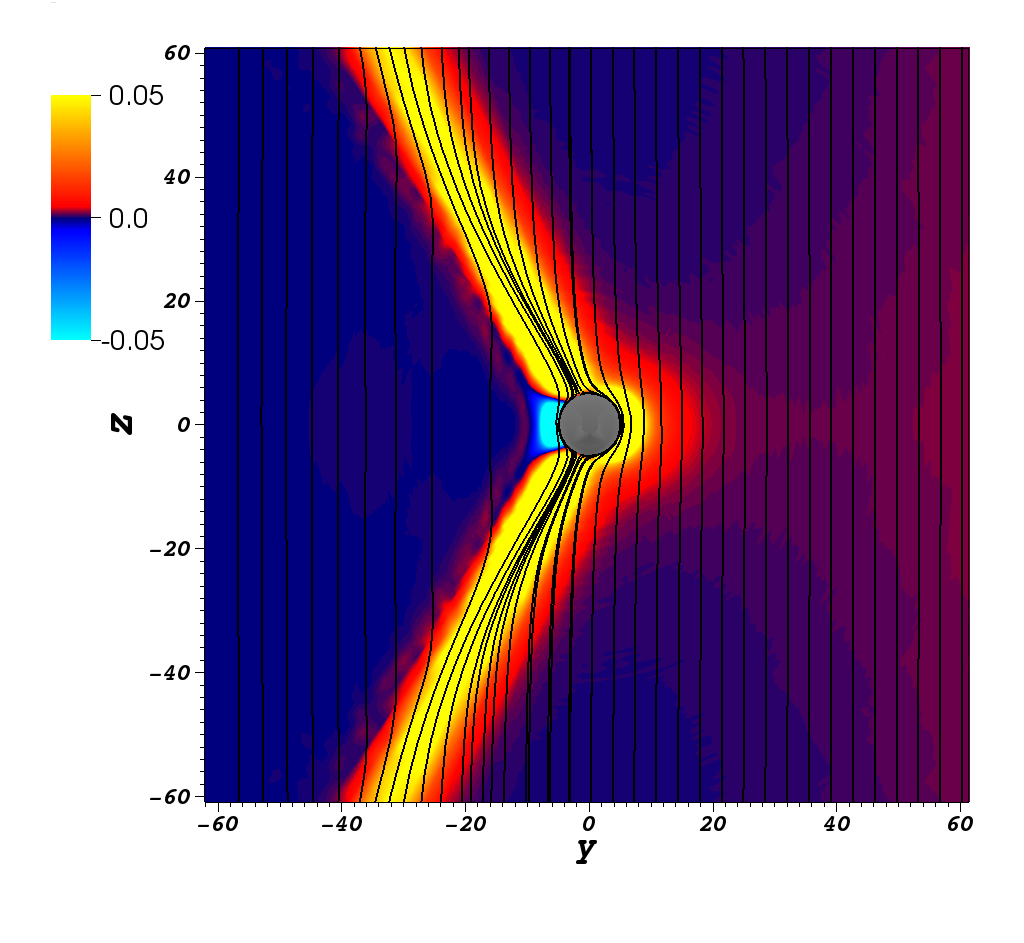}
\includegraphics[scale=0.23]{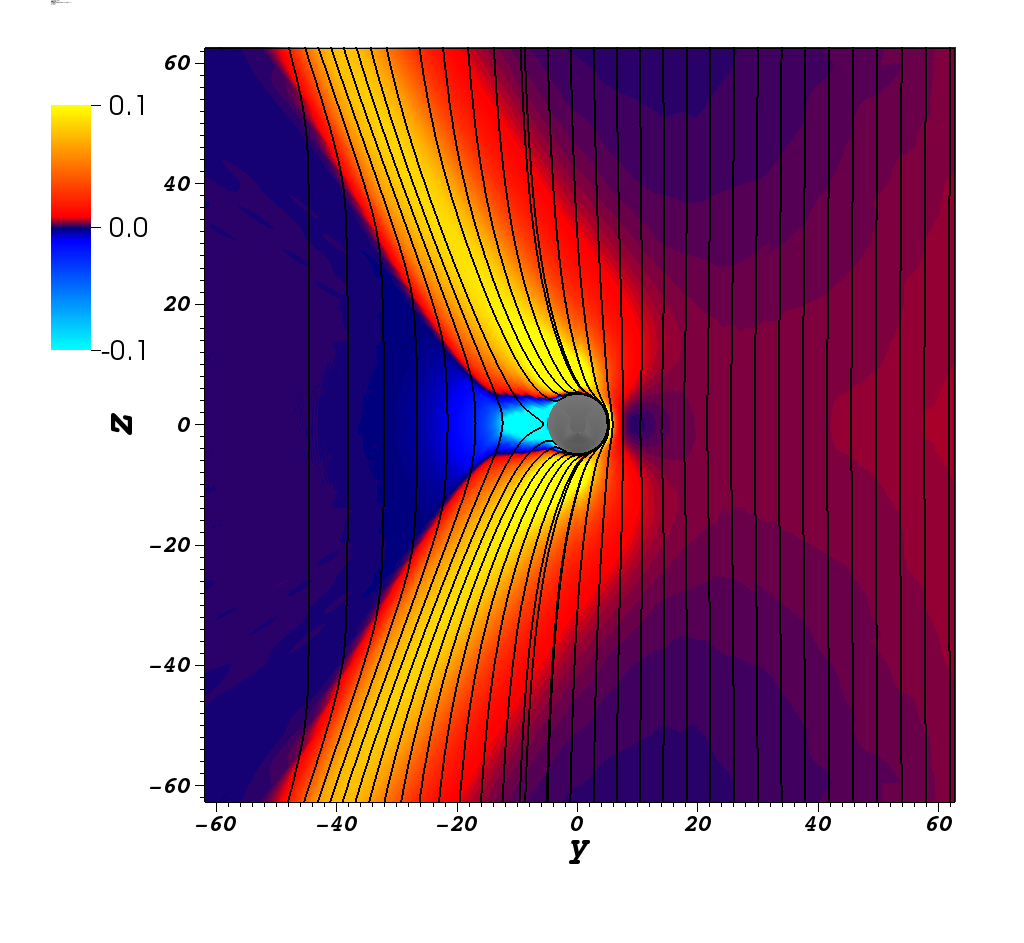}
  \caption{Late time numerical solution for a perfectly conducting sphere moving at speed $v=0.5$ along the $y$-axis. Magnetic fieldlines and flux density $p^{\prime a}N^{\prime}_{a}$ (in color scale) at the $x=0$ plane are represented in the top and bottom panels for stellar compactness of $\mathcal{C}=0$ and $\mathcal{C}=0.2$, respectively. The stellar surface is depicted by the gray disk in the center. }
 \label{fig:neutron_star}  
 \end{center}
\end{figure}
\begin{figure*}
	\centering
\includegraphics[scale=0.2]{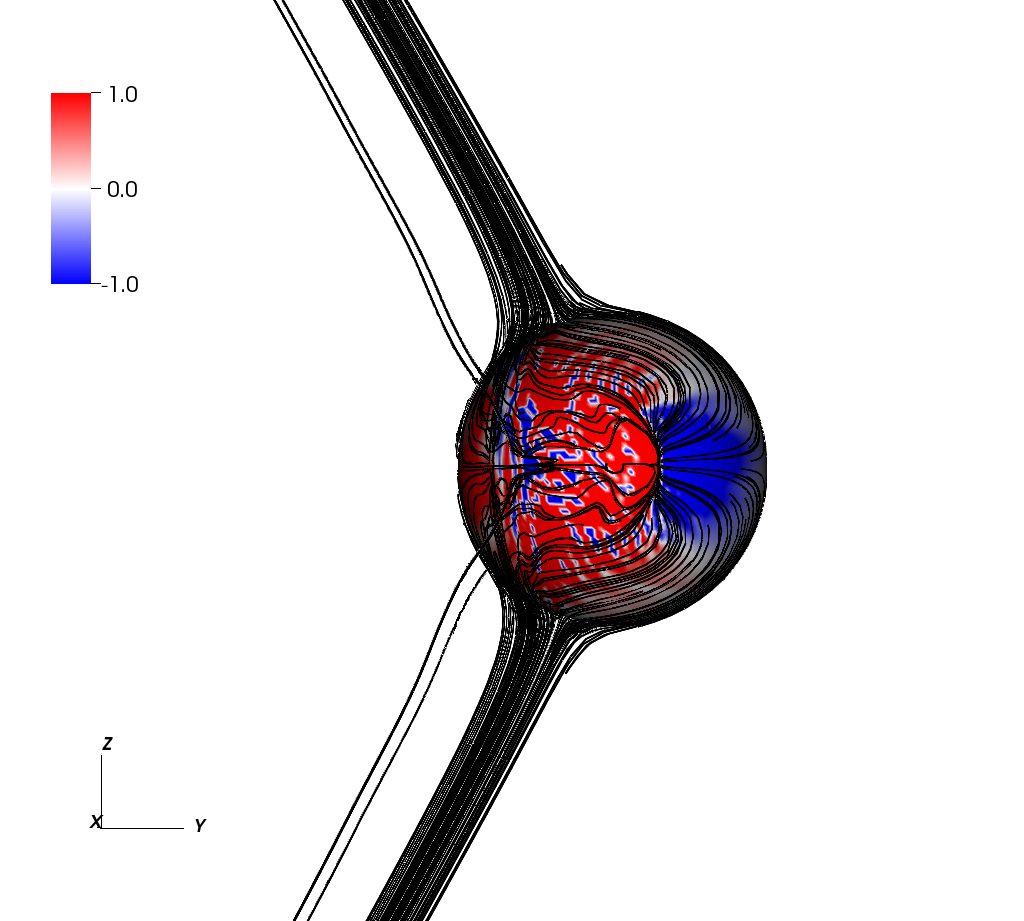}
\includegraphics[scale=0.16]{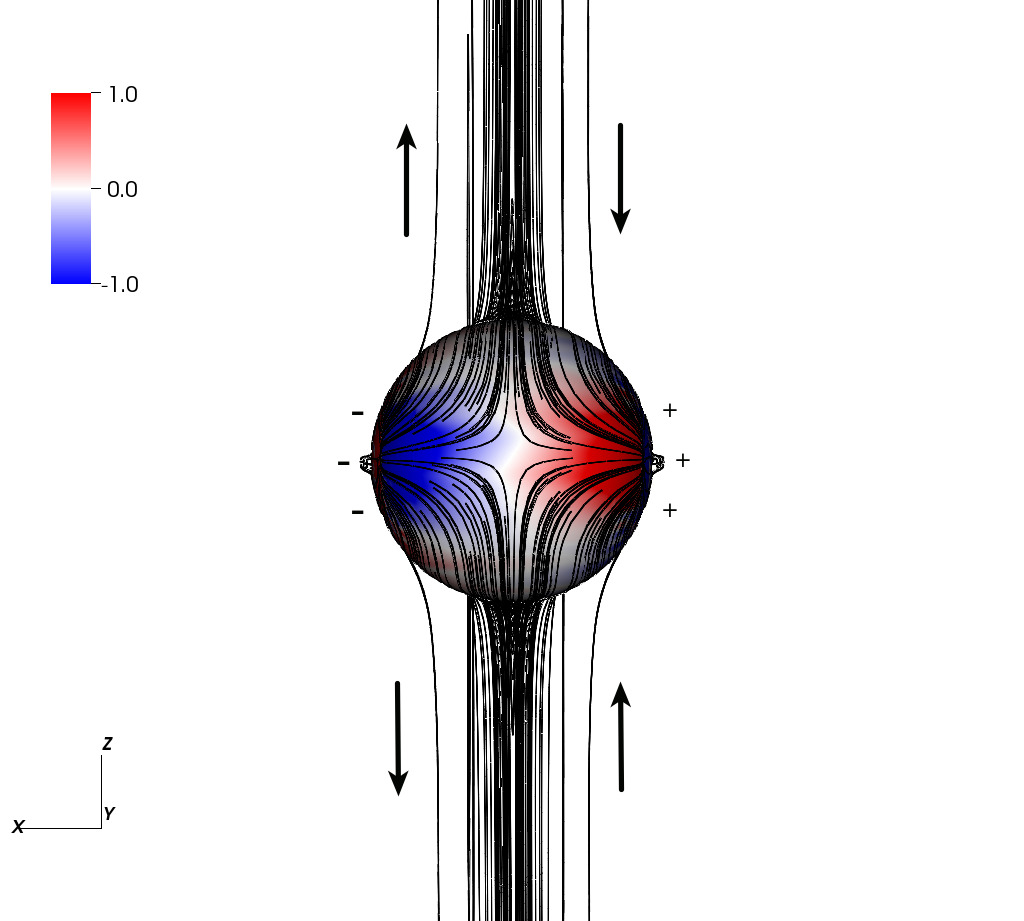}
\includegraphics[scale=0.18]{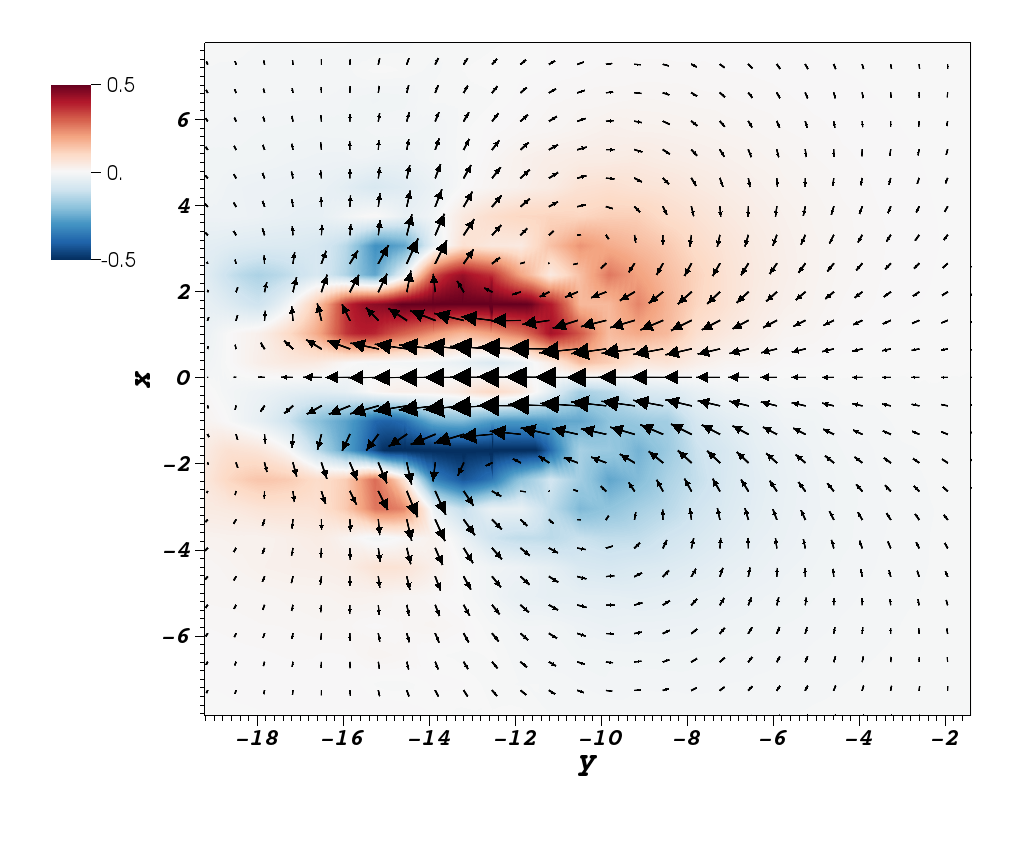}	
	\caption{\textit{Charge separation and induced electric circuit for the boosted NS in flat spacetime.} Streamlines of the currents are depicted, along with the charge density at the stellar surface (color scale): left and middle panels represents the side and front views, respectively. Right panel display the tangential magnetic field (arrows) and the normal component of the current density (in color scale) at the $z=20$ plane. 
	}
	\label{fig:circuit}
\end{figure*}
In Fig.\ref{fig:neutron_star} we show the Poynting fluxes produced by the neutron star moving at speed $v=0.5$, for the case discussed so far in flat spacetime (top panel), and also when including curvature effects setting a stellar compactness $\mathcal{C}=0.2$ (bottom panel). The overall qualitative picture is similar among these two cases, and also when comparing with the black hole scenario: there are positive electromagnetic fluxes along the (same) jet directions, where plasma currents are sustained by two counter-oriented twisted bundles of magnetic field lines (see right panel of Fig.~\ref{fig:circuit}). 
Even though the distortions on the magnetic field by the strong curvature of the BH are similar to the ones produced by the NS in flat spacetime, the underlying mechanism operating is quite different.  
As one might expect, the currents at the black hole horizon does not look like those at the conducting surface of the star. Moreover, there is no current sheet forming behind the star, as the one shown in Fig.~\ref{Current_sheet} for the BH, when there is no curvature (i.e.~$M=0$). But when the mass of the neutron star is tuned-on (i.e.~$M\neq0$), an analogous current sheet emerges and the emitted power gets enhanced. Thus, indicating that a composition of the two effects is acting; namely, the one associated with the perfect conductor condition and the one due to spacetime curvature.

\begin{figure}[H]
  \begin{center}
\includegraphics[scale=0.07]{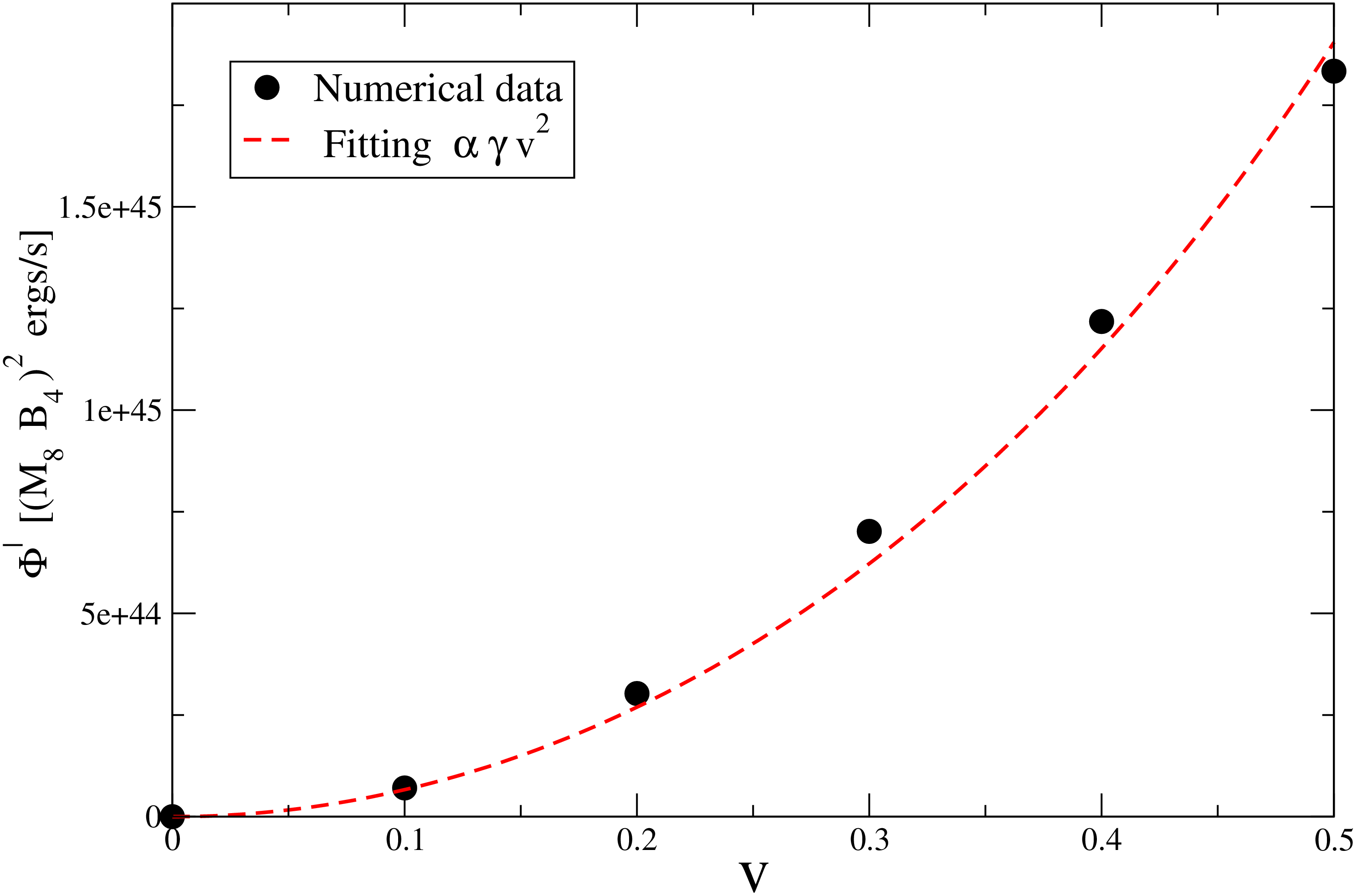}
  \caption{Dependence of the total collimated flux $\Phi^{\prime}$ on the speed $v$, for the late time solution of a NS of compactness $\mathcal{C}=0.2$. The red curve represents a fit of the form $\Phi$ $\propto$ $ \gamma v^{2}$.}
 \label{conducting_sphere} 
 \end{center}
\end{figure}

\begin{figure}
  \begin{center}
\includegraphics[scale=0.3]{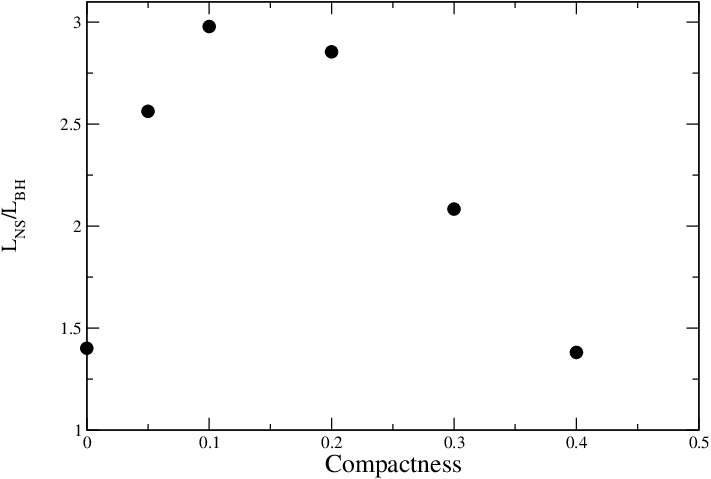}
  \caption{Luminosity produced by the NS, normalized by the one of the BH (i.e. $L_{NS}/L_{BH}$), as a function of the stellar compactness $\mathcal{C} \equiv M_{*}/R$ when moving at speed $v=0.3$ . 
}
 \label{compacticity} 
 \end{center}
\end{figure}

Quantitatively, the dependence of the collimated jet power on the boost speed is again $\propto \gamma v^2$ as in the black hole scenario (see Fig.~\ref{conducting_sphere}). 
Even though there is no unambiguous way to compare luminosities among a black hole and a perfectly conducting sphere in flat spacetime, we choose to relate the stellar radius $R$ with the BH mass $M$ by setting $R=2M$ in geometric units. This way, we will be comparing the plasma frame luminosity produced by the black hole with the one of a NS whose surface is placed at the Schwarzschild radius of the BH, exploring different stellar compactness at this fixed radius (Fig.~\ref{compacticity}). 
We find that the emitted power is now larger by a factor between $1.4$ and $3$, depending on the compactness. A similar enhanced luminosity for the NS scenario was found for rotating compact objects 
\cite{gralla2016electromagnetic}, with the explanation there being traced to distinct effective resistances of their electronic circuits. This argument, reminiscent from the membrane paradigm, indicates that the perfect conductor (placed at the BH horizon) in flat spacetime  would produce larger values ($\sim 40$\% in this case) simply because the black hole behaves effectively as a poorer conductor. We observe that increasing the stellar compactness then leads to an interesting interplay between the effects related with the conducting surface and those of gravity. The luminosity quickly rises (up to $\sim 3 L_{BH}$ by $\mathcal{C}=0.1$) and then smoothly begins to drop, presumably approaching the value $L_{BH}$ at compactness $\mathcal{C}\approx 0.5$. 

\subsection{Decomposing the solutions in physical modes}

\begin{figure*}
	\centering
\includegraphics[scale=0.16]{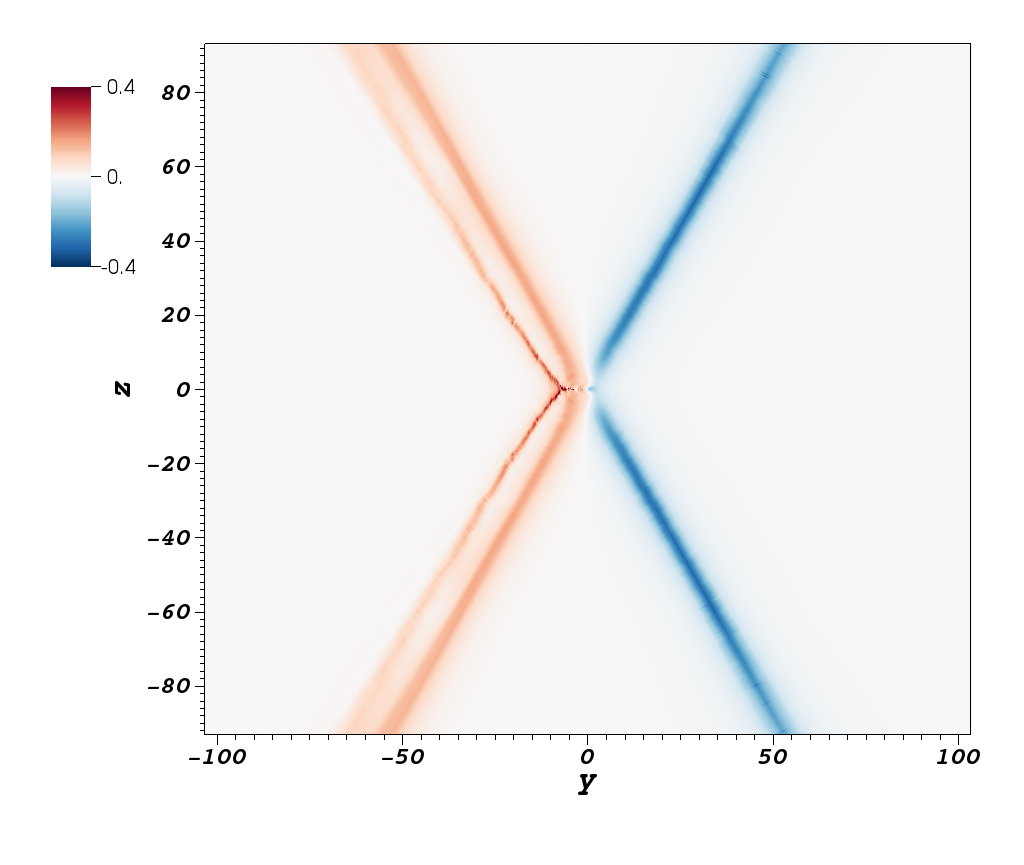}
\includegraphics[scale=0.16]{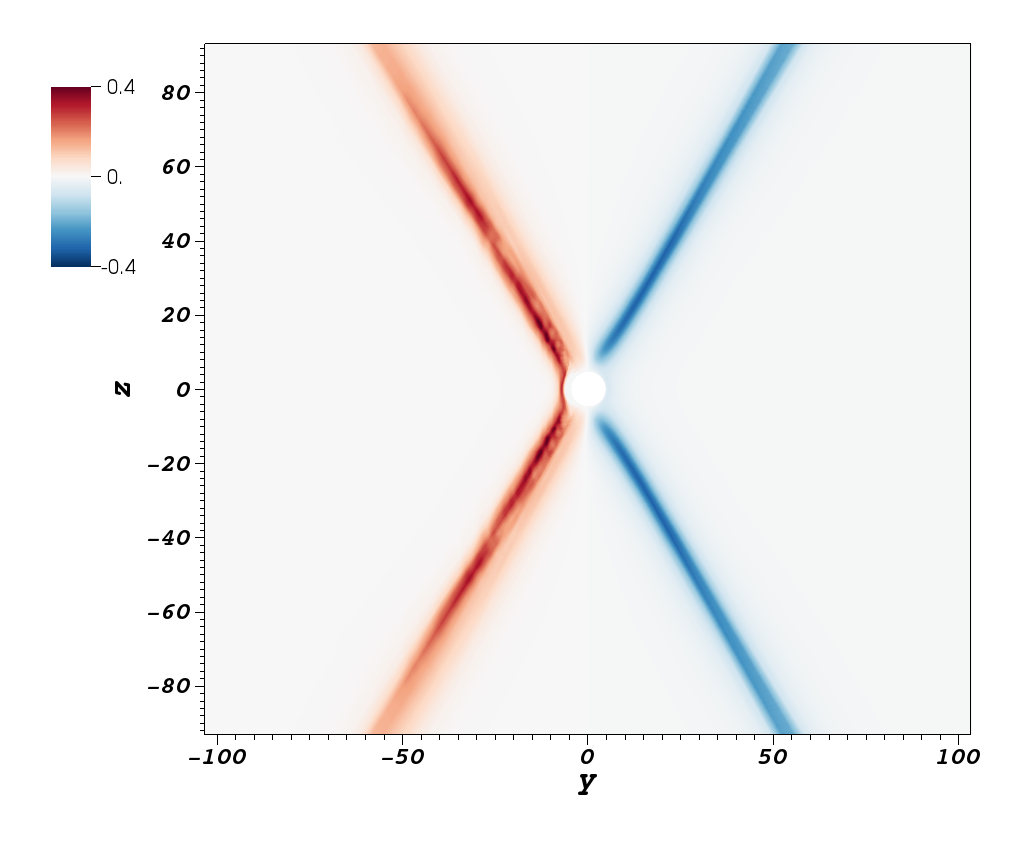}
\includegraphics[scale=0.16]{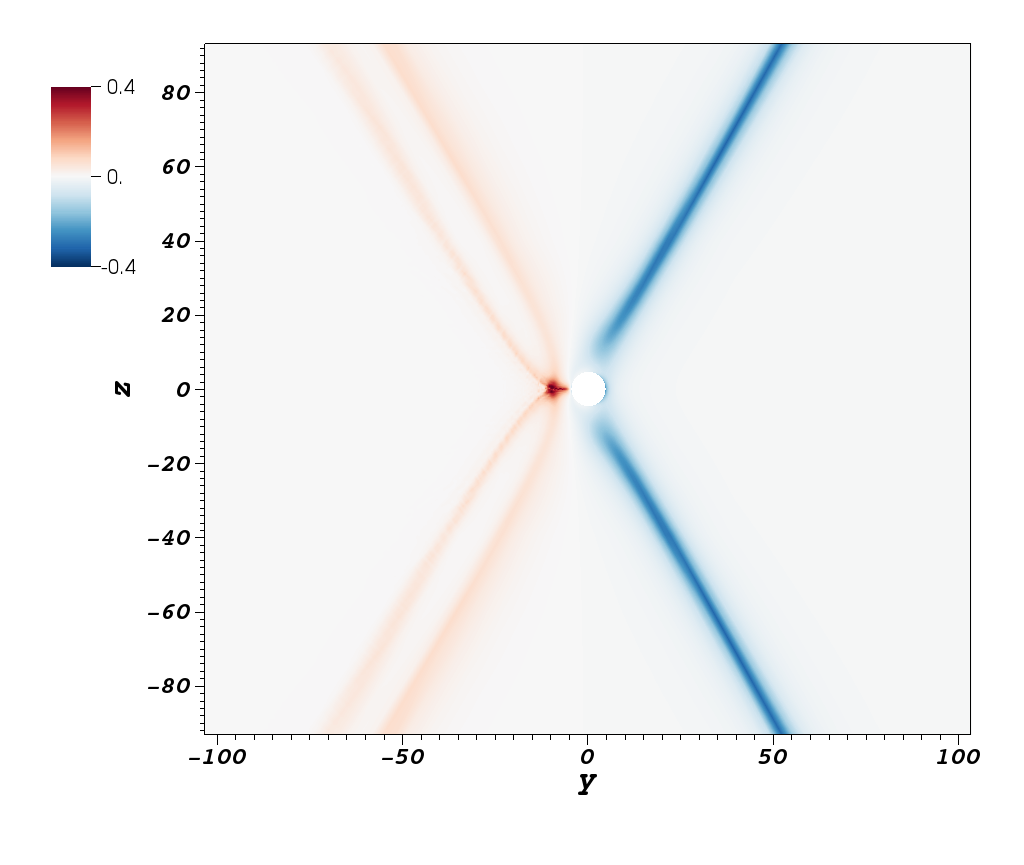}\\
\includegraphics[scale=0.16]{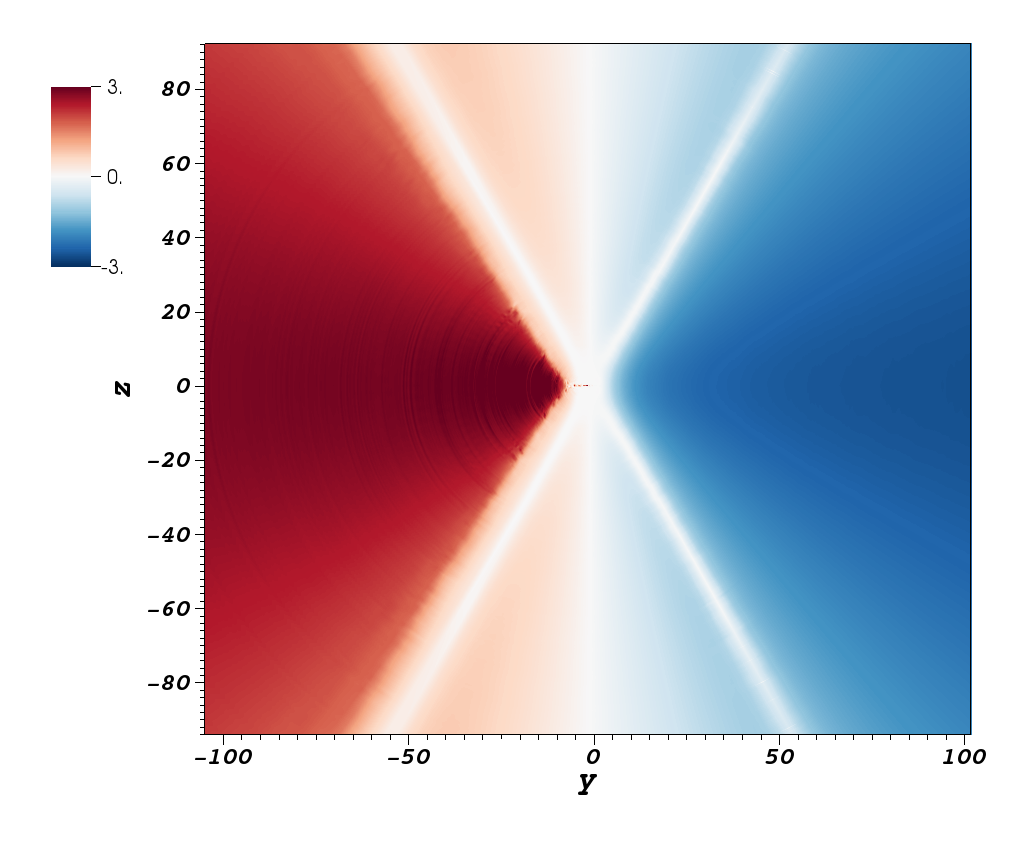}
\includegraphics[scale=0.16]{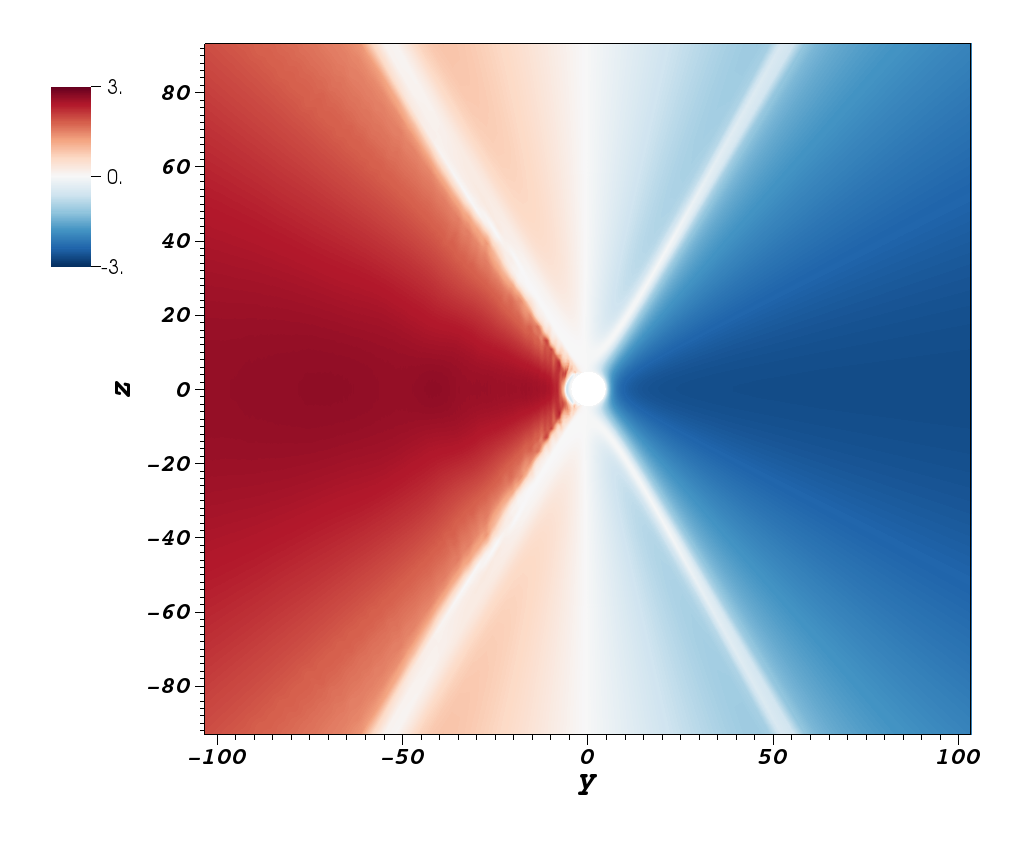}
\includegraphics[scale=0.16]{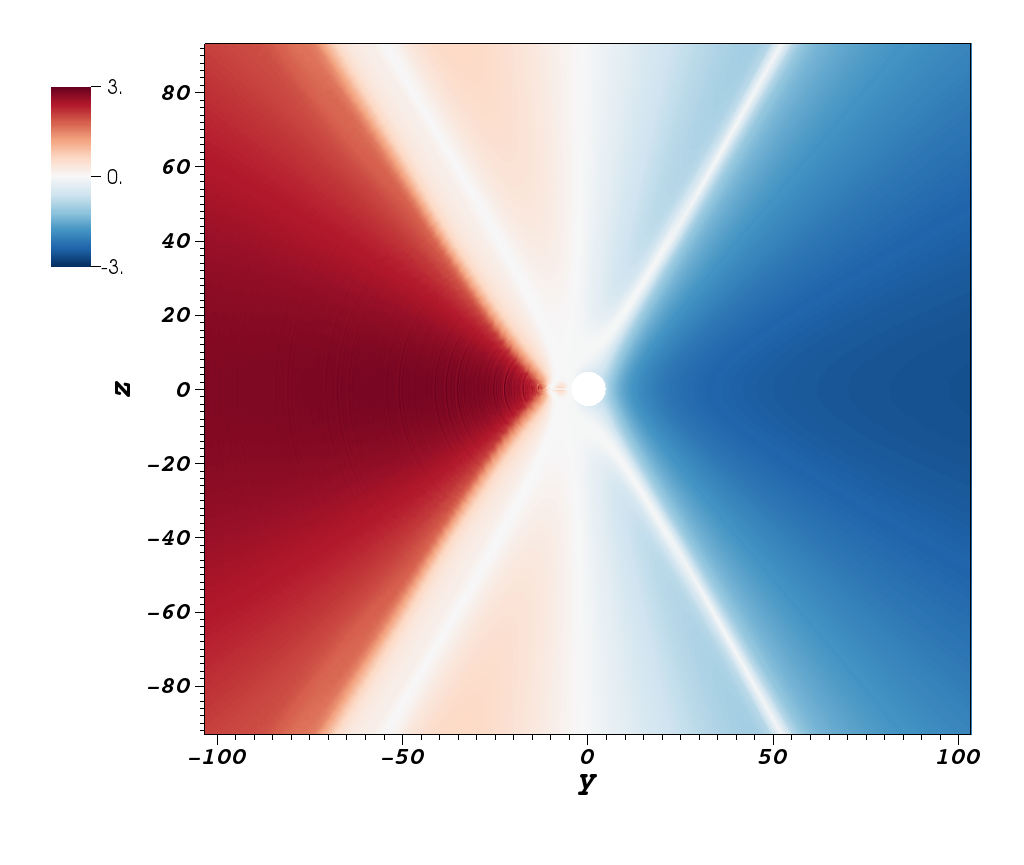}
	\caption{\textit{Net radial fluxes of physical modes at the $x=2$ plane}. Upper panels corresponds to the Alfvén modes, while lower panels show the fast magnetosonic modes. Compact objects moving at speed $v=0.5$ along the $y$-axis: black hole (left), neutron star in flat spacetime (middle) and neutron star of compactness $\mathcal{C}=0.2$ (right).  
	}
	\label{fig:modes}
\end{figure*}

Here we want to analyze the radial fluxes associated with the physical propagation modes, as defined by equations \eqref{A-flux} and \eqref{F-flux}. 
To that end, the fluxes are plotted in figure \ref{fig:modes}, for the three main cases under consideration. Namely, the black hole and the perfectly conducting sphere in flat/Schwazschild spacetime, moving at speed $v=0.5$ along the $y$-axis. 
Fast magnetosonic modes move at the speed of light without reference to the background magnetic field, thus here they seems to carry the contribution from the relative motion between the magnetized plasma and the object. On the other hand, Alfvén modes show funnels of positive radiation along the jets. But they also show similarly collimated regions of negative (i.e. incoming) flux at the opposite side of the $y=0$ plane. This is a bit puzzling, considering there is no electric charge density nor currents at those regions, as opposed to what happens with the positive components (outgoing) along the jets. It might be related, however, with the fact that we are computing these fluxes on the frame of the moving objects and not in the plasma frame; in that frame the net sum of the different contributions to the energy should be approximately zero when integrating on (e.g. spherical) surfaces enclosing the object.
%
\section{Discussion}

In this paper, we have studied jets arising from black holes and neutron stars moving across a magnetized force-free plasma environment. 
Even though, as one may expect, there is no conserved (Killing) energy being extracted from the moving objects --like e.g. emerging from the BH horizon--, we find however a positive (outgoing) net flux of approximate energy at the plasma frame. Such energy represents the one measured by an observer at rest with respect to the (asymptotic) uniform magnetic field.
We see that part of the available energy from the relative motion between the magnetized plasma and the object gets transferred to the electromagnetic field, producing two counter-oriented twisted bundles of magnetic fieldlines which induce stationary currents and support the highly collimated energy fluxes found; namely, the jets.  

By working on the co-moving reference frame, we were able to explore a wide range of boost velocities, finding a dependence of the luminosity of the form, $L \propto \gamma \, v^2$, in both scenarios.  Would there be an astrophysical situation where relative high velocities appear between compact objects and magnetic fields, the gamma factor would become preponderant.  We have also explored in some detail the other relevant parameters of the problem, like the orientation of the motion respect to the asymptotic magnetic field or the inclusion of black hole spin. We looked not only to the total energy flux, but also the contributions from each of the different force-free modes, namely Alven and fast modes. This way we were able to better understand the character of the process.

Comparing a black hole with a perfectly conducting sphere on flat spacetime, we have concluded that the overall effects are quite similar, although there are subtle but important differences among the two mechanisms. Clearly, the horizon does not behave as a perfect conductor\footnote{Notice that the spacetime might act as an effective conductor with finite resistivity, as suggested by the membrane paradigm approach (see e.g. Ref.~\cite{thorne}, and also \cite{penna2015} in the present context).} and, moreover, there is a strong current sheet emerging and playing an important role in the black hole case. 
We find that the perfect conductor generates about $40$\% larger luminosity than a black hole --when placed at the BH horizon--, in agreement with the familiar arguments from the membrane paradigm which states a BH would posses an effective finite conductivity.
Furthermore, we saw that when the mass of the neutron star is turned-on, a nontrivial superposition of these two mechanisms operates; interestingly, producing larger luminosities at intermediate values of the stellar compactness.

As mentioned before, we left the inclusion of the stellar magnetic field and rotation to a future work, where the interaction among this field and the external one can be explored in detail. This configurations could be used to mimic the presence of a NS companion orbiting in the context of a binary merger and to systematically study precursor electromagnetic signals along the lines of Refs.~\cite{palenzuela2013electromagnetic, ponce2014}.   


\section{Acknowledgments}

We acknowledge financial support from CONICET, SeCyT-UNC and MinCyT-Argentina.
F.C. acknowledge support from the Spanish Ministry of Economy and Competitiveness grants AYA2016-80289-P and AYA2017-82089-ERC (AEI/FEDER, UE).
This work used computational resources from Mendieta Cluster (Centro de Computación de Alto Desempeño, Universidad Nacional de C\'{o}rdoba), 
Pirayu Cluster (supported by the Agencia Santafesina de Ciencia, Tecnología e Innovación, Gobierno de la Provincia de Santa Fe, Proyecto AC-00010-18) and Centro de Cómputos de Alto Rendimiento (CeCAR). Which are all part of the Sistema Nacional de Computación de Alto Desempeño -- MinCyT-Argentina.


\bibliographystyle{unsrt} 
\bibliography{FFE}


\end{document}